\documentclass[10pt,final,journal,compsoc]{IEEEtran}
\usepackage[utf8]{inputenc}
\usepackage{adjustbox}
\usepackage[hidelinks]{hyperref}
\usepackage{multicol}
\usepackage{array,makecell}
\usepackage{rotating}
\usepackage{afterpage}
\usepackage{xcolor}
\usepackage{ragged2e}
\usepackage{arydshln}
\usepackage{graphicx}
\usepackage{wrapfig}
\usepackage[nobreak,space,compress]{cite}

\usepackage{multirow}
\usepackage{comment}
\usepackage{epsfig}
\usepackage{listings}
\usepackage{booktabs}
\usepackage{diagbox}
\usepackage{amsmath}
\usepackage{enumitem}
\setlist{leftmargin=0.692cm}

\newcommand{\ourapproach}{\mbox{LTM}} 
\usepackage[hang,flushmargin]{footmisc} 
\newcommand{\summarybox}[1]{
    \vspace{3mm}
    \noindent 
    \framebox[\linewidth][c]{\parbox[b]{0.95\linewidth}{{#1}}}
}
\makeatother

\begin{document}
\title{
{\ourapproach: Scalable and Black-box Similarity-based Test Suite Minimization based on Language Models
}}

\author{
    Rongqi Pan, 
    Taher A. Ghaleb, and
    Lionel C. Briand, \IEEEmembership{Fellow, IEEE}
\IEEEcompsocitemizethanks{
    \IEEEcompsocthanksitem R. Pan is with the School of EECS, University of Ottawa, Ottawa, Canada.~E-mail: rpan099@uottawa.ca

    \IEEEcompsocthanksitem T. A. Ghaleb is with the Computer Science Department, Trent University, Peterborough, Canada (Part of this
work was done while the author was with the University of Ottawa, Canada).~E-mail: taherghaleb@trentu.ca

    \IEEEcompsocthanksitem L. C. Briand holds shared appointments with the Lero SFI Centre for Software Research, University of Limerick, Ireland and the school of EECS, University of Ottawa, Ottawa, Canada.~E-mail: lbriand@uottawa.ca
}

\thanks{
This work was supported by a research grant from Huawei Technologies Canada Co., Ltd, as well as by the Mitacs Accelerate Program, the Science Foundation Ireland grant 13/RC/2094-2, and the Canada Research Chair and Discovery Grant programs of the Natural Sciences and Engineering Research Council of Canada (NSERC). The experiments conducted in this work were enabled in part by support provided by the Digital Research Alliance of Canada (https://alliancecan.ca).
}
}

\markboth{IEEE TRANSACTIONS ON SOFTWARE ENGINEERING}%
{Pan \MakeLowercase{\textit{et al.}}: \ourapproach: Scalable and Black-box Similarity-based Test Suite Minimization based on Language Models}

\IEEEtitleabstractindextext{
    \begin{abstract}\justifying
Test suites tend to grow when software evolves, making it often infeasible to execute all test cases with the allocated testing budgets, especially for large software systems. Test suite minimization (TSM) is employed to improve the efficiency of software testing by removing redundant test cases, thus reducing testing time and resources while maintaining the fault detection capability of the test suite. Most existing TSM approaches rely on code coverage (white-box) or model-based features, which are not always available to test engineers. Recent TSM approaches that rely only on test code (black-box) have been proposed, such as ATM and FAST-R. The former yields higher fault detection rates (\textit{FDR}) while the latter is faster. To address scalability while retaining a high \textit{FDR}, we propose \ourapproach~(\textbf{L}anguage model-based \textbf{T}est suite \textbf{M}inimization), a novel, scalable, and black-box similarity-based TSM approach based on large language models (LLMs), which is the first application of LLMs in the context of TSM. To support similarity measurement using test method embeddings, we investigate five different pre-trained language models: CodeBERT, GraphCodeBERT, UniXcoder, StarEncoder, and CodeLlama, on which we compute two similarity measures: Cosine Similarity and Euclidean Distance. Our goal is to find similarity measures that are not only computationally more efficient but can also better guide a Genetic Algorithm (GA), which is used to search for optimal minimized test suites, thus reducing the overall search time. 
Experimental results show that the best configuration of \ourapproach~(UniXcoder/Cosine) outperforms ATM in three aspects: (a) achieving a slightly greater saving rate of testing time ($41.72\%$ versus $41.02\%$, on average); (b) attaining a significantly higher fault detection rate ($0.84$ versus $0.81$, on average); and, most importantly, (c) minimizing test suites nearly five times faster on average, with higher gains for larger test suites and systems, thus achieving much higher scalability. 

    \end{abstract}

}
\maketitle
\IEEEdisplaynontitleabstractindextext
\IEEEpeerreviewmaketitle

\section{Introduction}

Software testing is an essential activity that ensures high-quality software systems by detecting faults in software releases~\cite{khan2018systematic}. 
When test suites are large, especially for large code bases, the scalability of running all test cases in a test suite quickly becomes a critical issue, as time and resources are limited for testing in industrial contexts~\cite{hemmati2013achieving}. The problem is particularly acute as test suites tend to increase in size over time with software evolution, making it impossible to run all test cases for every code change with allocated testing budgets~\cite{khan2018systematic}. To address this issue, test suite minimization (TSM) is widely employed to make the testing process more efficient while maintaining its fault detection capability~\cite{khan2018systematic}. In contrast to test case selection~\cite{yoo2012regression}, which selects a subset of test cases that are relevant to code changes, and test case prioritization~\cite{yoo2012regression}, which ranks test cases based on certain characteristics, such as fault detection capability, the objective of test suite minimization is to eliminate redundant or similar test cases that are unlikely to detect different faults~\cite{yoo2012regression,khan2018systematic}.

Though there exist various TSM approaches, most of them rely on either code coverage information (white-box)~\cite{yoo2012regression}, which requires access to software production code, or model-based features, such as transition coverage for UML state machine-based testing~\cite{hemmati2013achieving}. This information is not always accessible by test engineers and can be challenging to collect~\cite{pan2022test,arrieta2019pareto}. Further, collecting code coverage information is expensive~\cite{cruciani2019scalable,herzig2018testing} and can lead to scalability issues for large software systems.
Cruciani et al.~\cite{cruciani2019scalable} and Pan et al.~\cite{pan2023atm} addressed this issue by proposing TSM approaches that rely only on the source code of test cases (black-box). FAST-R, which is based on clustering algorithms, has been shown to be much more efficient than white-box approaches, but achieved relatively low fault detection rates. ATM, which is based on test case similarity and evolutionary search, achieved higher fault detection rates within practically acceptable, yet longer time. Since test suite minimization is performed on major software releases~\cite{noemmer2020evaluation,pan2023atm}, as opposed to test case selection and prioritization which are performed for every code change, ATM offers a better trade-off compared to FAST-R in many practical situations.
However, ATM still presents scalability issues for very large software systems, since its total minimization time (the sum of preparation time and search time) increases rapidly with test suite size, represented by the number of test cases per project version.
Our analysis of the total minimization time of ATM reveals that similarity measures play a major role in limiting its scalability, due to the fact that (1) computing tree-based similarity is expensive, taking up to $41.20\%$ of the total minimization time, and (2) similarity measures impact the search convergence and speed. This motivated us to investigate similarity measures that are both more efficient to calculate and more informative to guide the search.

Language models pre-trained on programming languages convert test code into vector-based embeddings. This enables vector-based similarity measurement, which facilitates implementation optimizations and results in much higher computational efficiency than tree-based similarity measurement that relies on traversing AST trees. Moreover, these language models are pre-trained on large source code corpora with various code understanding and generation tasks, thus generating embeddings that capture informative syntactic and contextual details from test code. This suggests that such embeddings with vector-based similarity might be potentially more informative than tree-based similarity in guiding evolutionary search.
Therefore, in this paper, we propose \ourapproach~(\textbf{L}anguage model-based \textbf{T}est suite \textbf{M}inimization), a scalable, black-box similarity-based TSM approach that is based on pre-trained Large Language Models (LLMs) and vector-based similarity measures, making it the first application of LLMs in the context of TSM, to the best of our knowledge. 
\ourapproach~uses the source code of test cases (Java test methods), without requiring any preprocessing, as input to five alternative pre-trained language models, namely \textit{CodeBERT}~\cite{feng2020codebert}, \textit{GraphCodeBERT}~\cite{guo2020graphcodebert},\textit{UniXcoder}~\cite{guo2022unixcoder}, StarEncoder~\cite{li2023starcoder}, and CodeLlama~\cite{roziere2023code}, in order to extract test method embeddings. 
Considering that test suite diversity was reported to have a positive correlation with fault detection~\cite{hemmati2013achieving,hemmati2010reducing,aghababaeyan2023deepgd}, \ourapproach~employs two similarity measures: \textit{Cosine similarity} and \textit{Euclidean distance}, to calculate the similarity between the extracted embeddings. Using the calculated similarity values, \ourapproach~employs a Genetic Algorithm (GA) to minimize test suites, which searches for the most optimal subset of a test suite, for a given testing budget.
We evaluated \ourapproach~with the same dataset used to evaluate ATM (\textsc{Defects4J}), followed a similar experimental design, and used the same evaluation metrics of effectiveness and efficiency: Fault Detection Rate (\textit{FDR}) and Minimization Time (\textit{MT}), respectively. In addition, considering the potential variation in test case execution times, assessing minimized test suites based on their number of test cases only might not always be accurate~\cite{khan2018systematic}. Therefore, we extended our evaluation of minimized test suites using an additional metric: Time Saving Rate (\textit{TSR}).
Moreover, we optimized GA by utilizing a more efficient data structure to accelerate fitness calculation and enhance memory usage, which led to a $190$-fold reduction in minimization time, without requiring any additional computation resources.
Then, we identified the best configuration of \ourapproach~by considering both effectiveness and efficiency, among all its alternatives, and compared it to the two best ATM configurations. 
Finally, we expanded our experiments by running \ourapproach~on a much larger project that ATM could not handle, to further assess the scalability of our approach.

Specifically, we address the following research questions.
\begin{itemize}
    \item \textit{RQ1: How does \ourapproach~perform for test suite minimization under different configurations?}
    \ourapproach~achieves high \textit{FDR} results (an overall average \textit{FDR} of $0.79$ across configurations) for a $50\%$ minimization budget (i.e., the percentage of test cases retained in the minimized test suite). The best configuration of \ourapproach~is UniXcoder using Cosine similarity when considering both effectiveness ($0.84$ \textit{FDR} on average) and efficiency ($0.82$ min on average), which also achieves a greater time saving rate (an average \textit{TSR} of $41.72\%$). For the large project, $Closure$, UniXcoder using Cosine Similarity takes only $17.80$ min in terms of \textit{MT} and achieves an \textit{FDR} of $0.79$, while saving $52.55\%$ of testing time.
    
    \item \textit{RQ2: How does \ourapproach~compare to ATM?}
    The best configuration of \ourapproach~outperforms ATM by achieving significantly better \textit{FDR} ($0.84$ versus $0.81$, on average), and more importantly, running much faster than ATM ($0.82$ min versus $4.06$ min, on average), in terms of both preparation time (up to two orders of magnitude faster) and search time (up to one order of magnitude faster). The latter is particularly important on typically large industrial systems and test suites where such differences practically matter. 
\end{itemize}

To summarize, the contributions of this work are as follows:
\begin{itemize}
    \item We propose a novel black-box TSM approach ({\ourapproach}) that relies, for the first time, on LLMs and two distance functions based on the generated embeddings. We investigate and conduct a comprehensive comparison among five recent language models with various model architectures, parameter sizes, inputs, and tasks used for pre-training.

    \item  We optimize the search process of {\ourapproach} by utilizing a more efficient data structure for fitness calculation, which in turn reduced the search time by $190$ folds.

    \item  We conduct a thorough comparative analysis of the efficiency and effectiveness, as well as the achieved saving in testing time, of black-box TSM approaches based on a large-scale test suite minimization experiments involving $17$ Java projects with $835$ versions, thus yielding valuable insights into the relative performance of alternatives. Overall, the experiments took around three months in calendar time corresponding to 6 years of computation on a cluster with $83,216$ available CPU cores.

    \item We analyze the minimization time of {\ourapproach} and show that it is much more scalable than the state-of-the-art (SOTA) approaches by running five times faster on average---with even higher gains for larger systems and test suites typically encountered in practice---while achieving significantly higher fault detection rates, as a result of using LLMs for embeddings.

\end{itemize}

The rest of this paper is organized as follows. Section~\ref{RelatedWork} reviews related work and motivates our research. Section~\ref{Approach} describes our test suite minimization approach. Section~\ref{Validation} validates our approach, reports the experimental design and results, and discusses the implications in practice. Section~\ref{Threats} discusses the validity threats to our results. Section~\ref{Conclusion} draws conclusions and suggests future work.

\vspace{2pt}
\section{Related Work}
\label{RelatedWork}

Test suite minimization (TSM) aims at improving the efficiency of software testing by removing redundant test cases, thus reducing testing time and resources while maintaining the effectiveness of the test suite (i.e., fault detection capability)~\cite{khan2018systematic}. There are various approaches that have been proposed to address TSM, including (a) greedy heuristics-based approaches~\cite{miranda2017scope,noemmer2020evaluation}, which select test cases iteratively based on code coverage information, (b) clustering-based approaches~\cite{coviello2018clustering,liu2011user}, which group test cases based on the similarity of their coverage information, and (c) search-based approaches~\cite{zhang2019uncertainty,hemmati2013achieving}, which employ evolutionary search to find an optimal subset of the test suite using model-based features. Most of these approaches utilize code coverage information (white-box)~\cite{yoo2012regression} (i.e., requires access to the system production code) or model-based features, which are not always available to test engineers~\cite{pan2023atm,arrieta2019pareto}. Moreover, collecting code coverage information can result in up to $30\%$ time overhead~\cite{cruciani2019scalable,herzig2018testing}, making it not scalable for large test suites and systems. 

Viggiato et~al.~\cite{viggiato2022identifying} investigate the similarity of test cases that are written in natural language. Though their objective is to identify redundancy among test cases, as opposed to TSM within a budget, their similarity measures can priori be further used for this purpose. They convert test cases into vector-based representations using five text embedding techniques (i.e., Word2Vec~\cite{mikolov2013distributed}, BERT~\cite{devlin2018bert}, Sentence-BERT~\cite{reimers2019sentence}, Universal Sentence Encoder~\cite{cer2018universal}, and TF-IDF (Term Frequency–Inverse Document Frequency)~\cite{joachims1997probabilistic}). They then identify similar test cases utilizing clustering algorithms to group similar test steps based on two different similarity metrics (Word Mover’s
Distance (WMD)~\cite{kusner2015word} and Cosine Similarity). The results show that their approach achieves an F-Score of $83.47\%$ for identifying similar test cases. In summary, different from our work, their focus is on identifying redundant test cases written in natural language, as opposed to code, and they do not aim to optimize a test suite within a budget.

Philip et al.~\cite{philip2019fastlane} proposed a black-box approach, called FastLane, which relies on test and commit logs containing historical information, to predict test case outcomes (i.e., \textit{pass} or \textit{fail}) using logistic regression to skip their execution. Using such information, FastLane reduces testing time and resources by running only a subset of test cases instead of the whole test suite. Results showed that FastLane reached $99.99\%$ in terms of test outcome accuracy and saved up to $18.04\%$ of testing time. However, it requires historical information about multiple runs of test cases, thus making FastLane inapplicable for new or recent test cases. Moreover, FastLane cannot be easily adapted to given minimization budgets (i.e., a predefined percentage of test cases to be executed). 
Arrieta et al.~\cite{arrieta2019pareto} relied on test case inputs and outputs for simulation-based testing and employed Non-Dominated Sorting Genetic Algorithm II (NSGA-II)~\cite{deb2002fast} to find an optimal subset of test cases as a minimized test suite. Chang et al.~\cite{chang2022putting} detected redundancy in test cases written in natural language using word embedding techniques and Cosine similarity. However, such information is not accessible in many contexts, thus rendering such black-box approaches not applicable.

Cruciani et al.~\cite{cruciani2019scalable} proposed a black-box approach, called FAST-R, that relies solely on the source code of test cases. FAST-R converts test code into vectors using a term frequency model~\cite{turney2010frequency} and then employs a random projection technique~\cite{johnson1984extensions} to reduce the dimensionality of vectors. Based on these vectors, clustering-based algorithms were performed with the centroids of clusters selected as a minimized test suite. Results showed that, compared to white-box approaches, FAST-R was much more efficient in terms of total minimization time while achieving comparable effectiveness in terms of fault detection capability. However, FAST-R achieved relatively lower fault detection capability for Java projects, with median fault detection rates ranging from $0.18$ to $0.22$ for minimization budgets ranging from $1\%$ to $30\%$. With such low effectiveness, FAST-R is therefore not a viable option in many practical contexts.

To achieve a better trade-off between the effectiveness and efficiency of test suite minimization, Pan et al.~\cite{pan2023atm} proposed a black-box test suite minimization approach, called ATM, which relies on test code similarity and evolutionary search. ATM preprocessed test code by removing the information that is irrelevant to the testing rationale and then converted it into Abstract Syntax Trees (ASTs). Then, four different tree-based similarity measures (i.e., top-down, bottom-up, combined, and tree edit distance) were used to calculate similarity values between these ASTs. Finally, evolutionary search (i.e., GA and NSGA-II) was employed using the similarity values as fitness to find, for a given test budget, an optimal subset of the test suite that contains diverse test cases. Results showed that the best configuration of ATM (i.e., GA with combined similarity) achieved a better trade-off in terms of effectiveness and efficiency of test suite minimization than FAST-R by achieving a significantly higher average fault detection rate ($+19\%$) while running within practically acceptable time ($1.2~hours$ on average), thus making it a better option in many contexts compared to FAST-R. 

However, ATM suffers from limitations regarding its scalability for very large projects in the dataset, such as $Time$, which consists of $30k$ lines of code in its most recent version with large test suites (nearly $4k$ test cases per version), as minimization took more than $10$ hours per project version as compared to FAST-R ($2.17$ seconds) and random minimization ($0.006$ seconds).
The similarity measures used by ATM had a direct impact on its scalability. First, the similarity calculation time took up to $41.20\%$ (using tree edit distance) of the overall minimization time of ATM, which is due to the fact that converting test code into ASTs and calculating tree-based similarity based on ASTs are both time- and resource-consuming. Second, the search time, which was influenced by the employed similarity measures and the number of test cases per version, increased rapidly with the test suite size. Interestingly, though tree edit distance took much longer to calculate than other ATM similarity measures, it enabled GA search to converge faster towards the termination criterion (a fitness improvement of less than $0.0025$ across generations). Although GA with combined similarity was identified as the best ATM configuration based on the average minimization time across projects ($1.2~hours$), it is important to note that it was $2.5~hours$ slower than GA with tree edit distance for the largest test suite with nearly $4k$ test cases. Therefore, when dealing with large test suites, adopting a more informative similarity measure that facilitates faster search convergence can result in a notable reduction in minimization time, regardless of the computational cost associated with the similarity calculation.

In contrast to the above TSM approaches, \ourapproach~relies on test code without requiring any preprocessing, and employs large pre-trained language models (i.e., CodeBERT, GraphCodeBERT, UniXcoder, StarEncoder, and CodeLlama) and commonly used vector-based similarity measures (i.e., Cosine similarity and Euclidean distance)~\cite{vijaymeena2016survey} to generate more informative similarity values between test case pairs. The goal is to both calculate similarity more efficiently and better guide the search algorithm (GA) to converge faster to a better fault detection capability. 

\section{\ourapproach: Language Model-based Test Suite Minimization}
\label{Approach}

This section describes our approach for test suite minimization, called \ourapproach, relying on language models to compute test code similarity and enable the use of evolutionary search for minimization. Our goal is to find a black-box solution that achieves a better trade-off, in terms of scalability and effectiveness, than the latest state-of-the-art (SOTA) approach, namely ATM~\cite{pan2023atm}. Despite the higher fault detection rates achieved by ATM, it has limitations in terms of scalability, which are due to the transformation of test cases into ASTs and the time-consuming calculation of tree-based similarity, making it difficult to apply for very large software systems. Similar to ATM, \ourapproach~is also a similarity-based approach operating under the assumption that there exists a positive correlation between test suite diversity and its fault detection capability~\cite{hemmati2013achieving}. We aim to find a similarity measure that is not only more computationally efficient but also provides better guidance for the search to converge faster to a higher fault detection capability and thus result in less search time.
To achieve this, we employed pre-trained language models to generate code embeddings and used them as the basis to measure the vector-based similarity between test case pairs.
The computation of vector-based similarity is much more efficient than tree-based similarity requiring tree traversal. 
Moreover, language models have the capability to capture patterns based on the syntax and semantics of test case code without requiring preprocessing or feature extraction, making them a more suitable alternative in our context.

\begin{figure*}[hbt!]
    \centering
    \includegraphics[width=1.00\textwidth]{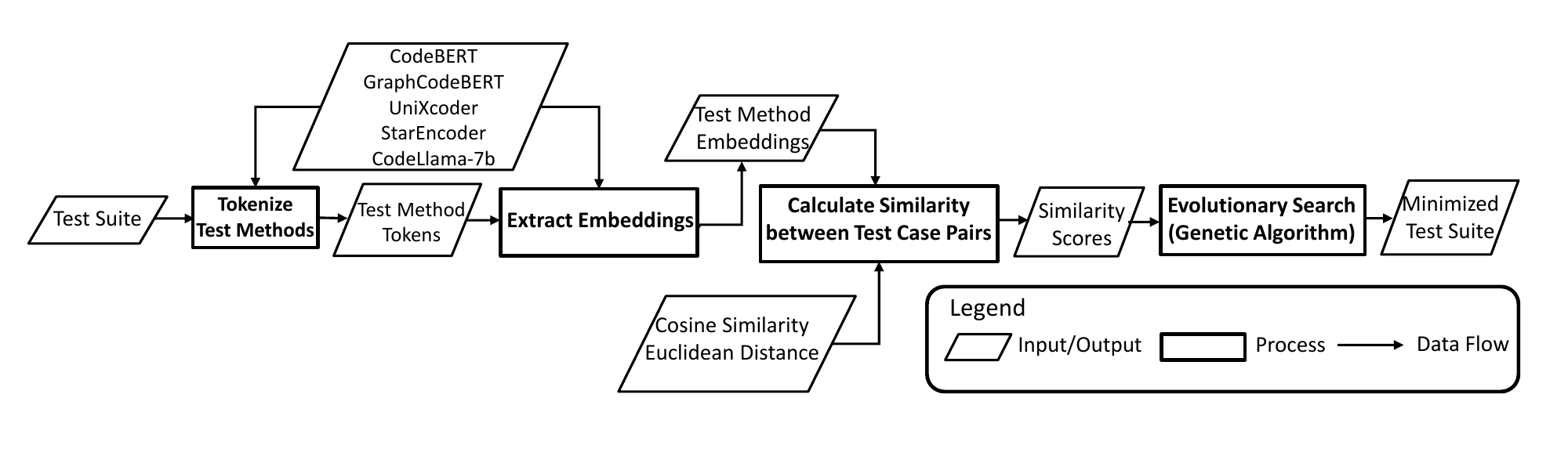}
    \vspace{-35pt}
    \caption{Main steps of \ourapproach~to perform test suite minimization}
    \label{fig:approach}
    \vspace{-3pt}
\end{figure*}
Figure~\ref{fig:approach} outlines the main steps of \ourapproach~to minimize test suites.
Regarding the first two steps, we describe how we tokenized the source code of test cases (Section~\ref{TestCodeTokenization}) and then extracted embeddings using language models (Section~\ref{TestCodeEmbeddings}). Then, we describe the algorithms (Cosine Similarity and Euclidean Distance) we employed for measuring the similarity between these embeddings (Section~\ref{SimilarityMeasures}). Finally, we describe the search algorithm (Genetic Algorithm) we employed to minimize test suites using the similarity measures as fitness. 

\vspace{2pt}
\subsection{Language Models for Test Method Representation}

ATM used ASTs (represented in XML format) to preserve the syntactic structure of test case code and calculated tree-based similarity between the ASTs of test case pairs, which we observed to be time- and resource-consuming due to the need for tree traversal, thus leading to scalability issues.
In addition, we found that among all tree-based similarity measures, ATM using tree edit distance achieved the highest \textit{FDR} and required the least search time across all project versions, since it made the search algorithm converge faster towards the termination criterion, that is fitness improvement is less than $0.0025$ across generations. This suggests that, regardless of the computational cost associated with its calculation, tree edit distance is a more informative similarity measure as it can be more effective in guiding the search to converge faster towards a higher fault detection rate. Therefore, we aim to develop a solution that overcomes the scalability issues of ATM by finding similarity measures that (1) do not require test code preprocessing,  (2) are computationally efficient, and (3) are more informative, offering better guidance to the search algorithm and accelerating its convergence, thus reducing search time.

In this paper, we employ five pre-trained language models, namely CodeBERT~\cite{feng2020codebert}, GraphCodeBERT~\cite{guo2020graphcodebert}, UniXcoder~\cite{guo2022unixcoder}, StarEncoder~\cite{li2023starcoder}, and CodeLlama~\cite{roziere2023code}. We select these models for the following reasons: 
\begin{itemize}
    \item CodeBERT, GraphCodeBERT, and UniXcoder are open-source publicly available language models that are specifically designed for programming languages and pre-trained on a large source code corpus. 
    These language models surpassed the performance of other SOTA models in code clone detection tasks~\cite{feng2020codebert, guo2020graphcodebert,guo2022unixcoder,lu2021codexglue}, relying on the similarity between pairs of code fragments\cite{guo2022unixcoder}, thus making them viable options for similarity-based test suite minimization.
    \item Similar to CodeBERT and its follow-up models, StarEncoder is a language model designed for understanding and analyzing code. Though it was not assessed for code clone detection tasks, StarEncoder is more recent and was pre-trained on a dataset encompassing more diverse ($86$) programming languages, enabling it to recognize test case patterns written in different coding styles.
    \item Unlike the above models that are pre-trained for code understanding tasks, CodeLlama is specialized for code generation and infilling tasks (i.e., filling a missing part of a code snippet). It has billions of parameters, which is significantly higher than the parameter size ($125 Million$) of the above models, and was pre-trained on a larger dataset, making it potentially able to distinguish test code more precisely.

\end{itemize}

In our context, these language models take the source code of test cases as input, without requiring any preprocessing or transformation into other formats, and generate embeddings that capture both semantic and contextual information from test case code. The output of these language models is represented as numeric vectors, which offer the opportunity to employ a variety of vector-based algorithms to measure similarities between test cases.
The process starts by passing the source code of a test case (a Java test method in our context) as input to the language models, which is then converted to a list of tokens, each of which is assigned an individual numerical vector representation. The final output is a vector representation (embedding) generated for the test method, which aggregates the information from all individual vector representations, which is called a test method embedding in our context.

\subsubsection{Tokenizing Test Methods}
\label{TestCodeTokenization}
Language models deal with a test code fragment (method) as a sequence of tokens and split it using byte-pair-encoding (BPE)~\cite{sennrich2015neural}, which is a subword segmentation algorithm that splits words into a sequence of sub-words. This enables language models to better handle out-of-vocabulary words, such as method and variable names. For example, the tokenizer splits the method name \textit{testCloning} into \textit{Ġtest} (where \textit{Ġ} denotes a space), \textit{Cl} and \textit{oning}, since the \textit{testCloning} word does not exist in the vocabulary and is thus separated into these three sub-words. The generated tokens are then processed as follows.

\begin{itemize}
    \item CodeBERT, GraphCodeBERT, and StarEncoder add two special tokens, namely $[CLS]$ and $[SEP]$, to the beginning and end of each sequence of tokens, respectively. The $CLS$ token is a special token that represents the whole input sequence, whereas the $[SEP]$ token is a separator token that denotes the end of the sequence~\cite{devlin2018bert}. In summary, each test method is represented as $[CLS],c_1,c_2,...,c_m,[SEP]$, where $c_i$ denotes the $i^{th}$ code token and $m$ is the total number of code tokens. 
    \item For UniXcoder, besides the $[CLS]$ and $[SEP]$ tokens, an additional special token ($[Enc]$, $[Dec]$, or $[E2D]$) is added to the beginning of the input indicating the pre-training mode of the model as encoder-only, decoder-only, or encoder-decoder mode, respectively. These modes differ in terms of model architectures and training tasks during pre-training, thus supporting various downstream tasks. We used the encoder-only mode for producing contextualized code embeddings, as decoder-related modes are used for code generation~\cite{guo2022unixcoder}. Therefore, the final input for UniXcoder is represented as $[CLS],[Enc],[SEP],c_1,c_2,...,c_m,[SEP]$.
    \item CodeLlama uses $<$$s$$>$ and $<$$/s$$>$ tokens denoting the beginning and end of each token sequence, respectively:  ($<$$s$$>$$,c_1,c_2,...,c_m,$$<$$/s$$>$).
\end{itemize}

During pre-training, each token is then mapped to a $768$-dimensional vector representation that contains the semantic and contextual information of this token for all language models above, except for CodeLlama where the vector size is $4,096$. We set the token length to $512$ for all language models during the tokenization process.

\subsubsection{Generating Test Method Embeddings}
\label{TestCodeEmbeddings}
~
A test method embedding is a numerical vector representation of a test method that captures semantic and contextual information from the source code. It is based on how and on what data a model was pre-trained, as follows.
\begin{itemize}
    \item CodeBERT pre-trained code representations are based on a large corpus called CodeSearchNet~\cite{husain2019codesearchnet}, which contains both natural language (e.g., code comments) and source code across six programming languages. Its follow-up models (GraphCodeBERT and UniXcoder) leveraged data flow information during pre-training, which captures relationships between variables in the input code fragments and ASTs of the source code, respectively, to enhance the code representation.
    \item StarEncoder was pre-trained on a large dataset encompassing 86 programming languages collected from The Stack~\cite{kocetkov2022stack}, which is specifically crawled from GitHub repositories for pre-training code LLMs.
    \item CodeLlama was pre-trained on a massive dataset ($500 Billion$ tokens) containing both source code and natural language. CodeLlama has multiple variations and model sizes, depending on the type of training data and parameter size. In this paper, considering the importance of scalability, we used the smallest version of CodeLlama (i.e., CodeLlama-7b), which has seven billion parameters.
    
\end{itemize}
In terms of model architecture:

\begin{itemize}
    \item CodeBERT and StarEncoder employ a multi-layer bidirectional self-attentive Transformer~\cite{vaswani2017attention} as model architecture to help the model capture contextual and positional information for each token from the entire input sequence.
    
    \item GraphCodeBERT extends such architecture using a graph-guided masked attention function to help the model encode graph-based data flow information.
    
    \item UniXcoder, on the other hand, is based on a multi-layer Transformer with a prefix denoting the pre-training mode of the model. The model architecture for the encoder-only mode, which is employed by \ourapproach, allows the model to learn, for each token, the contextual information from the entire input sequence. 
    \item CodeLlama uses an optimized auto-regressive transformer, enabling the model to efficiently predict missing parts within code sequences. Despite being a decoder-only model, CodeLlama exhibits the capability to understand the test code context and generate missing parts accurately. 
    
\end{itemize}

\noindent During pre-training:

\begin{itemize}
    \item CodeBERT employs Masked Language Modeling (MLM)~\cite{devlin2018bert}, which allows the model to predict the masked tokens from the input sequence and whether a token was randomly replaced in a given sequence~\cite{clark2020electra}. These two tasks help the model generate code embeddings that contain more accurate contextual information, accounting for the position of the tokens in the source code input.
    \item GraphCodeBERT also employs MLM with two additional tasks: Edge Prediction~\cite{guo2020graphcodebert} and Node Alignment~\cite{guo2020graphcodebert}. Edge Prediction allows the model to predict which edges, which denote dependencies between variables, have been masked for a given variable in the data flow graph. Node Alignment allows the model to predict which code token is related to a given variable in the data flow graph. These two tasks further enhance the code representation using data flow information.
    \item UniXcoder also uses the MLM task for the encoder-only mode, and further learns the test method embedding, which is first generated by taking the average (i.e., mean pooling) of all token embeddings and then further trained by utilizing multi-modal contrastive learning~\cite{gao2021simcse} and cross-modal generation~\cite{guo2022unixcoder}, which both leverage AST and code comments to further enhance test method embeddings. Multi-modal contrastive learning generates different embeddings for the same input but with different dropout masks, which allows the model to predict the original embedding using the other generated embeddings. This task helps the model to better distinguish between different method embeddings and thus can better deal with downstream tasks, such as test method similarity analysis~\cite{gao2021simcse}. Cross-modal generation generates code comments for the input test method. This helps the model to fuse the semantic information from the code comments, which describe the function of the code, to the test method embeddings. 
    \item StarEncoder also employs MLM in addition to Next Sentence Prediction (NSP)~\cite{devlin2018bert}, which helps the model learn the relationship between test code segments by predicting whether segments follow other segments.

    \item Unlike the above models, CodeLlama employs the code infilling task, which allows the model to generate the missing part of a code snippet while comprehending the entire context. This enables the model to understand the logical structure of the test code and the relationships between test code segments.

\end{itemize}

For \ourapproach~using CodeBERT, GraphCodeBERT, and StarEncoder, we use the embedding that corresponds to the $[CLS]$ token as a test method embedding, since it aggregates the information from all tokens of that method. For \ourapproach~using UniXcoder, we use the pre-trained test method embedding corresponding to each test method. For \ourapproach~using CodeLlama, we rely on the mean pooling of the last hidden states (i.e., the average of all token embeddings extracted from the last hidden layer of the model), which is also an effective strategy that aggregates the information from all code tokens~\cite{ma2019universal}.

\subsection{Similarity Measurement of Test Method Embeddings}
\label{SimilarityMeasures}

\ourapproach~employs two similarity measures for calculating the similarity between test method embeddings: Cosine Similarity and Euclidean Distance. They measure the similarity between test method embeddings from different aspects: Cosine similarity measures the angle between two vectors, whereas Euclidean distance calculates the straight-line distance between them.\\

\noindent\textbf{Cosine similarity.} This is a measure of similarity between two vectors based on the cosine of the angle between them~\cite{gomaa2013survey}, which is the dot product of the vectors divided by the product of their lengths, and is calculated as follows:

    \begin{equation}
     Cosine~Similarity = \frac{\mathbf{T_1} \cdot \mathbf{T_2}}{\|\mathbf{T_1}\|\|\mathbf{T_2}\|}
    \end{equation}
where $\mathbf{T_1}$ and $\mathbf{T_2}$ denote the embeddings of test case $T_1$ and $T_2$, respectively. 

The value of Cosine similarity ranges from $-1$ to $1$. The higher the value of Cosine similarity, the more similar the two test cases are. In order to bound the value of Cosine similarity between $0$ and $1$, we normalized it as follows: 

    \begin{equation}
     Norm.~Cosine~Sim. = 1 -  \frac{\arccos{\frac{\mathbf{T_1} \cdot \mathbf{T_2}}{\|\mathbf{T_1}\|\|\mathbf{T_2}\|}}}{\pi}
    \end{equation}

\noindent\textbf{Euclidean distance.} This is a measure of similarity between two vectors based on the square root of the sum of squared differences between corresponding elements of the two vectors~\cite{gomaa2013survey}, which is calculated as follows:

    \begin{equation}
     Euclidean~Distance = (\sum_{i}^{m}{(t_{1i} - t_{2_i})^2})^{1/2}
    \end{equation}
where $t_{1i}$ and $t_{2i}$ is the $i^{th}$ element of embedding $T_1$ and $T_2$, respectively. $m$ is the total number of elements in each embedding ($4,096$ for CodeLlama and $768$ for the other models).

The value of Euclidean Distance ranges from $0$ to $\infty$. The higher the value of Euclidean distance, the less similar the two test cases.
In order to bound the Euclidean distance between $0$ and $1$, and obtain a similarity measure, we normalized the distance as follows~\cite{deza2009encyclopedia}:

    \begin{equation}
     Norm.~Euclidean~Dist. = \frac{1}{1 + (\sum_{i}^{m}{(t_{1i} - t_{2_i})^2})^{1/2}}
    \end{equation}

We used the `\texttt{pdist}' function\footnote{\url{https://docs.scipy.org/doc/scipy/reference/generated/scipy.spatial.distance.pdist.html}} from the \textit{Scipy} Python library, which performs pairwise calculations for large datasets. The `\texttt{pdist}' employs highly optimized C code to improve the efficiency of similarity calculation on vector-based data, and stores the output in a condensed matrix, which further reduces the required memory resources.

\subsection{Search-based Test Suite Minimization}
\label{SearchBased}

Given that test suite minimization is an NP-hard problem~\cite{hemmati2013achieving}, it can be addressed efficiently using meta-heuristic search algorithms~\cite{pan2023atm}  to find feasible, near-optimal solutions, a minimized test suite containing diverse test cases for a given budget in our context. Like ATM, we relied on a Genetic Algorithm (GA) since it has shown, in the context of TSM, to achieve a better trade-off between effectiveness and efficiency than its multi-objective alternative, namely Non-Dominated Sorting Genetic Algorithm II (NSGA-II)~\cite{pan2023atm}.
The optimization problem addressed by the GA is defined as a fixed-size subset selection problem~\cite{pymoo}. Each solution (chromosome) is a subset and is represented as a binary vector where $1$ denotes the selection of the test case and $0$ otherwise. The vector length equals the total number of test cases in the test suite before minimization. Given specified minimization budgets (25\%, 50\%, and 75\%), the percentage of selected test cases in each solution is set to equal to the budget. For the fitness function, we used the summation of the maximum squared similarity values (i.e., Normalized Cosine Similarity and Normalized Euclidean Distance), shown below:

\begin{align}
\hspace{-6pt}
   Fitness = \frac{\sum_{i,t_i\in M_n}(Max_{i,j, t_i,t_j\in M_n, i \neq j}\ Sim(t_i,t_j))^2}{n}
\end{align}
where $M_n$ is a minimized test suite of size $n$, and $Sim(t_i,t_j)$ is the normalized similarity score of test cases pair $t_i$ and $t_j$.

To select hyperparameter values, we followed the same published guidelines\footnote{\url{https://www.obitko.com/tutorials/genetic-algorithms/recommendations.php}} as ATM for GA search, by using a population size of $100$, a mutation rate of $0.01$, and a crossover rate of $0.90$. The GA process maximizes the diversity of the subset by iteratively evolving the population using crossover and mutation operators while evaluating each subset using the predefined fitness function. For each generation, the size of each solution (subset) remains fixed and equal to the minimization budget. This process is repeated until the fitness value improvement across generations is less than $0.0025$. 

\begin{figure}[hbt!]
    \centering
    \includegraphics[width=0.28\textwidth]{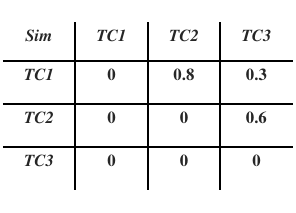}
    \vspace{-12pt}
    \caption{An example of how similarity values of pairs of test cases are represented using a matrix. Values on and below the diagonal are set to 0 as they are either useless or duplicates.}
    \label{fig:sim_matrice}
    \vspace{-3pt}
\end{figure}

In terms of implementation, \ourapproach~optimizes GA search by changing the data structure of the input (i.e., similarity values) to (1) accelerate the fitness calculations and (2) enhance memory usage. The data structure employed by ATM for storing similarity values is a data frame~\footnote{https://pandas.pydata.org/docs/reference/api/pandas.DataFrame.html}, which has a complex indexing and labeling mechanism, thus making fitness calculation computationally expensive. Therefore, we use matrices~\footnote{https://numpy.org/doc/stable/reference/generated/numpy.array.html} instead, since they facilitate faster numerical calculations due to the fact that (1) it is a simpler data structure without complex indexing and labeling and (2) the matrix operations are implemented using highly optimized C code that ensures efficient computation and memory management.
As shown in Figure~\ref{fig:sim_matrice}, we stored the similarity values for all pairs of test cases in a matrix format without any duplicate and useless values (the similarity values on and below the diagonal of the matrix were set to 0). To save memory and further accelerate calculations, we employed a sparse matrix format~\footnote{https://docs.scipy.org/doc/scipy/reference/sparse.html}, which exclusively preserves the value and position (indicating for which pair of test cases a given similarity value represents) of non-zero similarity values. 
Note that we did not parallelize the fitness calculation as it significantly increases the memory required by the matrix, which must contain similarity values for all considered subsets, making it less scalable. Although fitness calculations can be parallelized in other ways, such as multiprocessing, which utilizes multiple CPUs to parallelize fitness calculation, or distributed computing~\cite{pymoo}, which distributes the fitness calculation to various devices or nodes, they require additional computational resources, which contradicts our goal of improving scalability.
         
\section{Validation}
\label{Validation}
\vspace{-2pt}

This section reports on the experiments we conducted to assess the effectiveness and efficiency of \ourapproach.

\subsection{Research Questions}

    \noindent\textbf{\textit{RQ1: How does \ourapproach~perform for test suite minimization under different configurations?}}
    \vspace{2pt}
        Test suite minimization aims to reduce the number of test cases in a test suite by removing redundant test cases while maintaining its fault detection power, thus achieving higher testing efficiency and effectiveness. In particular, the performance of \ourapproach~can be influenced by (a) the language models used for extracting test method embeddings and (b) the distance functions used for measuring the similarity between test method embeddings. In this RQ, we assess the performance, in terms of effectiveness and efficiency, of \ourapproach~under different configurations, each with a different combination of language model and distance function. We evaluated \ourapproach~using three minimization budgets (i.e., $25\%$, $50\%$, and $75\%$).

        \begin{itemize}
        \item \textbf{\textit{RQ1.1: How effectively can \ourapproach~minimize test suites?}}
        The language models used for \ourapproach~(i.e., CodeBERT, GraphCodeBERT, UniXcoder, StarEncoder, and CodeLlama-7b) vary in terms of architectures and training data, thus generating embeddings that capture different information about test case code. The distance functions we used to measure the similarity or distance between test method embeddings differently, which can lead to different comparison results. Therefore, the combination of language models and distance functions can affect the ability of \ourapproach~to remove redundant test cases, which can in turn affect its fault detection capability. In this RQ, we assess the effectiveness of \ourapproach~in terms of fault detection capability under different configurations. In addition, we calculated the execution time of minimized test suites to quantify the achieved time savings resulting from test suite minimization.
        
        \vspace{4pt}
        \item \textbf{\textit{RQ1.2: How efficiently can \ourapproach~minimize test suites?}}
        Test suite minimization should run within practical time. The use of different combinations of language models and distance functions can impact the preparation time, and the resulting similarity values can guide the GA search to varying degrees of efficiency, thus resulting in different search time. In this RQ, we assess the efficiency of \ourapproach~in terms of preparation time (taken for test code tokenization, embeddings extraction, and similarity calculation) and search time (taken for GA search).
    \end{itemize}

    \vspace{3pt}
    \noindent\textbf{\textit{RQ2: How does \ourapproach~compare to ATM?}}
    \vspace{2pt}
    ATM was developed as a black-box test suite minimization approach that achieved a better trade-off, in terms of effectiveness and efficiency, than the SOTA approach (FAST-R~\cite{cruciani2019scalable}). However, the minimization time of ATM increased rapidly with test suite size, making it less scalable for large projects with large test suites. For example, ATM required more than $10$ hours, in terms of total minimization time, to minimize a large project in the dataset used for its evaluation, thus indicating scalability issues. In this RQ, we assess the performance of \ourapproach~compared to the two best configurations of ATM, specifically ATM using GA with combined and tree edit distance similarity measures, in terms of both effectiveness and efficiency. 
    
        \begin{itemize}
        \item \textbf{\textit{RQ2.1: How effectively can \ourapproach~minimize test suites compared to ATM?}}
        We aim to find an informative similarity measure that can better capture similarities between test cases, thus achieving higher \textit{FDR}. In this RQ, we compare the best configuration of \ourapproach~to the two best configurations of ATM in terms of fault detection capability.

        \vspace{4pt}
        \item \textbf{\textit{RQ2.2: How efficiently can \ourapproach~minimize test suites compared to ATM?}}
        To achieve better scalability compared to ATM, we aim to reduce (a) preparation time, by investigating a similarity measure that is more efficient to calculate, and (b) search time, by investigating a more informative similarity measure that can better guide the search, thus resulting in faster convergence. In this RQ, we compare the efficiency of \ourapproach~to the two best configurations of ATM in terms of both preparation time and search time. Further, to better assess scalability for both \ourapproach~and ATM, we fit a regression model to investigate how minimization time (the sum of preparation time and search time) increases with the size of a test suite.

        \end{itemize}

\subsection{Experimental Design and Dataset}
  We conducted experiments to evaluate the performance of alternative \ourapproach~configurations, each with a different combination of language model and similarity measure, resulting in ten configurations in total (five different language models and two similarity measures), using the same experimental design and dataset as ATM.
  We conducted our experiments on a cluster of $1,340$ nodes with $83,216$ available CPU cores, each with a $2x$ AMD Rome 7532 with 2.40 GHz CPU, 256M cache L3, 249GB RAM, running CentOS 7.
  We considered each Java project version as an independent subject and conducted our experiments on all Java project versions ($835$ in total) in the dataset to increase the number of instances for experimental evaluation and, therefore, enhance the soundness of our conclusions. Each TSM approach was performed on every project version and ran $10$ times to account for randomness and draw statistically valid conclusions, using three minimization budgets ($25\%$, $50\%$, and $75\%$), which is the percentage of test cases that retained in the minimized test suite). There is a total of $25,050$ runs ($835$ Java project versions $\times$ $10$ runs per version $\times$ $3$ minimization budgets) for each configuration. Overall, the experiments, including similarity calculations, GA search, as well as executing all test suites $10$ times,  took more than three months in calendar time corresponding to 6 years of computation. 
  We identified the best configuration of {\ourapproach} considering both the effectiveness and efficiency and compared it to the two best configurations of ATM. 
  Our data, scripts, and raw results can be found in our replication package~\cite{our_replication_package}

\subsubsection{Baseline approach (ATM)}
\label{sec:baseline}

    We compared \ourapproach~to ATM~\cite{pan2023atm}, a similarity-based, search-based test suite minimization approach that relies solely on the source code of test cases. 
    ATM preprocesses test code by removing information that is irrelevant to testing logic from the test code, then converts test code into ASTs. ATM calculates four different tree-based similarity measurements between test case ASTs: top-down, bottom-up, combined and tree edit distance similarity. Note that ATM does not use the full code (i.e., the source code of test cases without preprocessing) as it would make the already-long preparation phase of ATM much longer and impractical in realistic settings. This is due to the fact that (1) the XML files used by ATM to store Abstract Syntax Trees (ASTs) of test cases would require substantial storage space and (2) it would take longer time to traverse the ASTs of test cases when calculating similarity, thus yielding much longer preparation time for transforming the test case code into ASTs and calculating similarity values. As a result, ATM's preparation time can take up to 41.2\% of the total minimization time, and using the full code in ATM would make matters much worse. 
    
    After the preprocessing step, search algorithms (i.e., GA and NSGA-II) were then employed based on similarity values to find an optimal subset of test cases by removing redundant test cases. 
    Using GA with combined similarity, ATM achieved a better trade-off in terms of effectiveness and efficiency compared to a SOTA test suite minimization approach, FAST-R~\cite{cruciani2019scalable}. However, it suffered from scalability issues as, for example, it took more than $10$ hours to minimize $4k$ test cases for the largest test suite. This motivated us to compare \ourapproach~to ATM, in terms of fault detection capability and, more importantly, minimization time. We aim to assess whether \ourapproach~can be more scalable than ATM while achieving comparable or even higher fault detection capability.
    We compared \ourapproach~to the two best configurations of ATM, using GA with combined similarity (ATM/Combined) and tree edit distance (ATM/TreeEditDistance). The former was the best configuration considering the trade-off between effectiveness and efficiency, whereas the latter achieved higher fault detection capability but was more time-consuming than the former in terms of total minimization time. We further evaluated the achieved saving in testing time for every minimized test suite for both {\ourapproach} and ATM. We relied on the publicly available replication package of ATM\footnote{\url{https://doi.org/10.5281/zenodo.7455766}} provided by its authors.
    
\subsubsection{Dataset} 

   To facilitate comparisons, we ran our experiments using the same dataset as ATM to evaluate the effectiveness and, more importantly, the scalability of~\ourapproach. The dataset consists of $16$ Java projects collected from \textsc{Defects4J}, a well-known dataset that offers \textit{real} and reproducible faults extracted from real-world, open-source Java projects to support software testing research. There is no publicly available industrial system or open source system that contains information to automatically trace test failures to system faults, a key requirement for our experiments.
   To further validate the performance of \ourapproach, we included an additional large project from Defects4J, called $Closure$, which was not part of the dataset used to evaluate ATM due to scalability issues.
   Each project has numerous faulty versions, each of which contains a single \textit{real} fault that is present in the production code and leads to the failures of one or more test cases. It is worth noting that there is currently no publicly available dataset that offers multiple \textit{real} faults per version, as automatically establishing a clear link between faults and test case failures would be challenging.
   Table\ref{tab:Subjects} presents the characteristics of all $17$ projects. Overall, the project sizes in terms of KLoc range from $2$ to $179$ KLoc, with the corresponding tests sizes ranging from $4$ to $253$ KLoc. The number of faulty versions ranges from $4$ to $174$ across projects, with the average number of test cases per version ranging from $152$ to $7,308$ across projects. Project sizes and tests sizes (i.e., lines of code) were extracted by analyzing the most recent version of each project using the \textit{CLOC}\footnote{\url{https://github.com/AlDanial/cloc}} tool.
   In summary, we evaluated {\ourapproach} using a total of $835$ versions from $17$ projects. However, given the scalability limitations of ATM~\cite{pan2023atm}, we compared our results to ATM using only the $661$ versions from the original $16$ projects, excluding the $Closure$ project.

\begin{table}[hbt!]
\centering
\vspace{-5pt}
\caption{Summary of the $17$ Java projects}
\vspace{-12pt}
 \resizebox{0.5\textwidth}{!}{
    \begin{tabular}{lcccc}
    \toprule
       \multirow{2}{*}{Project} & Project Size &  \# of versions     & Average \# of test cases & Tests Size\\
                               & (KLoC)~~~ &  (faults)~~~~  & (methods) per version~ & (KLoC)~~\\
    \midrule
       Chart           &  92    &  26  & 1,817  &   41   \\
       Cli             &  2     &  39  & 256    &   4   \\
       Codec           &  9    &  18  & 413    &   15   \\
       Collections     &  30    &  4   & 1,040  &   38   \\
       Compress        &  45    &  47  & 404    &   29   \\
       Csv             &  2     &  16  & 193    &   7   \\
       Gson            &  9     &  18  & 984    &   20   \\
       JacksonCore     &  31    &  26  & 356    &   45   \\ 
       JacksonDatabind &  74    &  112 & 1,814  &   72   \\
       JacksonXml      &  6     &  6   & 152    &   10   \\
       Jsoup           &  14    &  93  & 494    &   13   \\
       JxPath          &  20    &  22  & 250    &   6    \\
       Lang            &  30    &  64  & 1,796  &   61   \\  
       Math            &  71    &  106 & 2,078  &   73   \\
       Mockito         &  21    &  38  & 1,182  &   36   \\
       Time            &  30    &  26  & 3,918  &   56   \\
       Closure         &  179   &  174 & 7,308   &   253  \\
    \bottomrule
    \end{tabular}
}
\label{tab:Subjects}
\vspace{-5pt}
\end{table}

\subsubsection{Evaluation metrics}
We used the same evaluation metrics as ATM, specifically fault detection rate and total minimization time, in addition to the achieved saving in testing time of minimized test suites.

    \vspace{2pt}
    \noindent\textbf{\textit{Fault Detection Rate (\textit{FDR}).}} We use \textit{FDR} to assess the effectiveness of \ourapproach~compared to ATM. Given that each project version contains a single fault, the corresponding fault detection rate per version can either be $1$ (fault detected) or $0$ (fault undetected). Therefore, we calculate the fault detection rate for each project by considering all its versions, as follows:

    \begin{equation}
    FDR = \frac{\sum_{i = 1}^{m}f_i}{m}
    \end{equation}
    where $m$ denotes the total number of versions of the project. For a project version $i$, $f_i$ is equal to $1$ if at least one failing test case is included in the minimized test suite, indicating the detection of the fault of that version, or $0$~otherwise.

    For example, in the $Chart$ project, which has a total of $26$ faulty versions, if the minimized test suites detected the faults in 21 versions but did not in $5$ versions, \textit{FDR} would be calculated as $21/26$, resulting in an \textit{FDR} of $0.81$. Note that we ran our experiments for each version a total of $10$ times to mitigate randomness. Hence, the final \textit{FDR} for each project is obtained by taking the average \textit{FDR} results across the $10$ runs.

    \vspace{4pt}
    \noindent\textbf{\textit{Total Minimization Time (\textit{MT}).}} We assess the efficiency of \ourapproach~compared to ATM by computing (1) the \textit{preparation time}, which includes the time required for loading the language models, tokenizing test case code, extracting test method embeddings, and calculating similarity values between all pairs of test cases, and (2) the \textit{search time}, which is the time required for running the GA.

    \vspace{4pt}
    \noindent\textbf{\textit{Time Saving Rate (\textit{TSR}).}} We use \textit{TSR} to quantify the achieved testing time savings resulting from test suite minimization. It is important to note that the reduction in test execution time may not be directly proportional to the reduction in the test suite size, as the time required to execute individual test methods may vary~\cite{khan2018systematic}. Using the \textsc{Defects4J} infrastructure, we ran the test suite for each project version ($835$ versions in total) $10$ times and collected the average execution time for each test method in a clean, stable environment, a cluster with many available CPU cores. This ensures that each run runs independently and is not impacted by external interferences and fluctuations in memory or CPU performance. Then, we calculate the \textit{TSR} for each minimized test suite as follows:

    \begin{equation}
    TSR = (1 - \frac{test~execution~time~after~TSM}{test~execution~time~before~TSM})*100
    \end{equation}

    \vspace{4pt}
    \noindent\textbf{\textit{Fisher's exact test.}} We use Fisher's exact test~\cite{raymond1995exact}, a non-parametric statistical hypothesis test, to assess whether the difference in proportions of detected faults between the alternative \ourapproach~and ATM configurations is significant.

\subsection{Results}

In this section, given space constraints, we only report results for a minimization budget of 50\% (the percentage of test cases retained in the minimized test suite) as trends and conclusions are similar for other budgets (25\% and 75\%), which are available in our replication package~\cite{our_replication_package}.

\subsubsection{Using full test code versus preprocessed test code in {\ourapproach}}
\label{PreliminaryStudy}

One important question is whether the source code of test cases requires preprocessing, such as removing test case names, logging statements, comments, and test assertions. Though such preprocessing was performed by ATM~\cite{pan2023atm}, due to necessity (Section~\ref{sec:baseline}), this information may still convey relevant information about test case similarity, thus retaining them can potentially improve fault detection capability.
This information can also help language models better learn about semantic and contextual information about test cases and thus generate more informative test method embeddings. For example, code comment \textit{`Tests the equals method'} describes the functionality of the test code, thus providing potentially important semantic information~\cite{guo2022unixcoder}.
Therefore, we conducted a preliminary study to compare the \textit{FDR} results of LTM using full code (i.e., the source code of test cases without preprocessing) and preprocessed code on a sample of $10$ smaller projects from the \textsc{Defects4J} dataset. We observed that \ourapproach~achieved higher \textit{FDR} results using full code than using prepossessed code ($+0.02$ on average) across all configurations, thus making the former a better choice for our experiments. We summarize the results and provide further details in our replication package~\cite{our_replication_package}.

\vspace{2pt}
\subsubsection{Improvement using optimized GA.}
As discussed in Section~\ref{SearchBased}, we optimized GA to accelerate the search process. Our results revealed that, compared to \textit{MT} before optimization, the search time of ATM/Combined and ATM/TreeEditDistance is $224.66$ and $203.02$ times faster per project version, respectively. For \ourapproach, the search time is $190.48$ times faster per project version, across all configurations, which is a significant improvement in scalability. Note that the following comparisons between \ourapproach~and ATM are based on applying this optimized GA to both approaches.

\subsubsection{RQ1 results}

Tables~\ref{tab:FDR_results}, \ref{tab:MT_results}, and \ref{tab:TSR_results} provide the results of \ourapproach~in terms of \textit{FDR}, \textit{MT} (in min), and \textit{TSR} (in percentage), respectively, when using CodeBERT, GraphCodeBERT, UniXcoder, StarEncoder, and CodeLlama-7b with Cosine and Euclidean similarity measures, for a $50\%$ minimization budget.

\vspace{4pt}
\noindent\textbf{\textit{RQ1.1 results.}}
    Table~\ref{tab:FDR_results} shows that all \ourapproach~configurations achieved a high \textit{FDR}, ranging from $0.65$ to $0.95$ across projects, with both the mean and median ranging from $0.75$ to $0.84$ for all configurations, for a $50\%$ minimization budget.
    The highest overall average \textit{FDR} was achieved using UniXcoder/Cosine (mean~=~$0.84$, median~=~$0.82$), ranging from $0.75$ to $0.95$ across projects.
    However, \ourapproach~using CodeBERT/Euclidean also yielded a high \textit{FDR} (mean~=~$0.82$, median~=~$0.84$).
    CodeBERT/Euclidean and UniXcoder/Cosine showed standard deviations of $0.06$ and $0.07$, respectively.
    In addition, Fisher's exact test results reveal that \ourapproach~using UniXcoder/Cosine achieved significantly better \textit{FDR} results than all other configurations, with $\alpha = 0.01$.
    
\begin{figure}[ht]
    \centering
    \includegraphics[width=1.0\columnwidth]{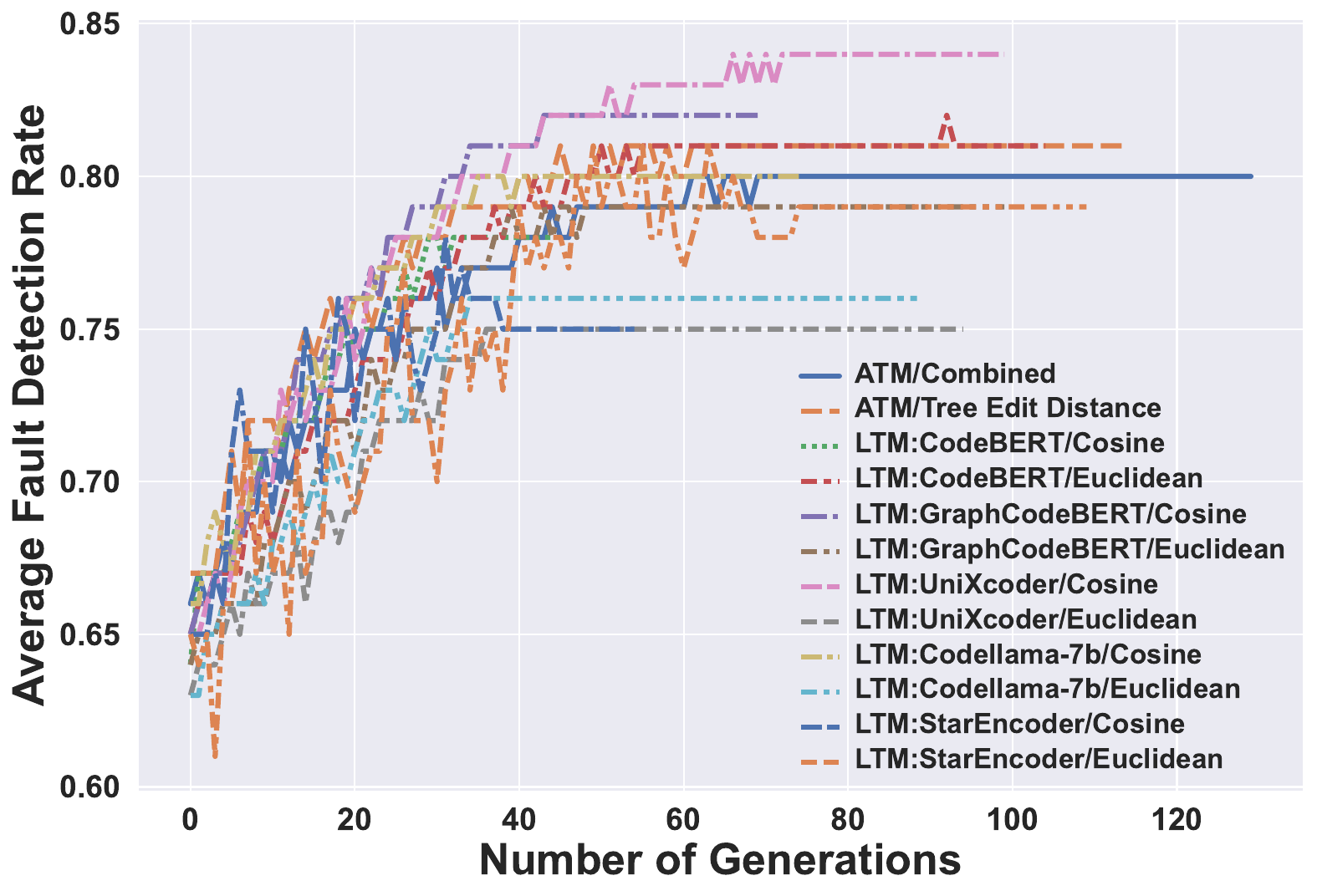}
   
    \caption{\textit{FDR} across projects for each generation of \ourapproach~and ATM}
    \vspace{-5pt}
    \label{fig:convergence_plot}
\end{figure}

    Figure~\ref{fig:convergence_plot} depicts the average \textit{FDR} for all projects across generations, for all configurations of \ourapproach. Compared to other configurations, \ourapproach~using UniXcoder/Cosine converges faster to a higher \textit{FDR} ($0.84$ across projects), suggesting that it offers better guidance for the GA search, making it more effective at removing redundant test cases and thus leading to a better fault detection rate. 

   We observed that, for each language model, there seems to be a significant difference in \textit{FDR} between using Cosine Similarity and Euclidean Distance, which is further confirmed by Fisher's exact test with $\alpha = 0.01$. This suggests that \textit{FDR} results are influenced by the combination of language models and distance functions. 
   The embeddings generated by UniXcoder and GraphCodeBERT are enriched with detailed information about the nodes and edges within test code ASTs. This could be the reason that Cosine Similarity, as a directional similarity, is more informative in distinguishing such test method embeddings than Euclidean Distance, a magnitude similarity.
   For CodeLlama-7b, the embedding dimensionality is $4,096$, compared to $768$ for other language models, which can explain its better performance with Cosine Similarity over Euclidean Distance, since the latter is less effective in high-dimensional spaces~\cite{xia2015effectiveness}.

    In certain projects ($Cli$, $Codec$, $Csv$, and $Time$), UniXcoder/Cosine yielded a $5\%$ to $10\%$ lower \textit{FDR} than other configurations. However, Fisher's Exact Test revealed that those differences are not significant. Nevertheless, we provide some tentative explanations regarding these projects:  
    (1) A larger training data set covering a greater variety of programming languages (StarEncoder) and a substantially greater number of parameters (CodeLlama) might be beneficial for understanding test code from these projects, and (2) using ASTs as input for pre-training might introduce unnecessary details into test method embeddings, if test cases in these projects are similar in terms of code structure but rather differ regarding low-level details (e.g., complex string manipulation or domain-specific character strings), thus making UniXcoder less effective in understanding their test code.

    In addition, we observed that UniXcoder/Cosine achieved a higher average \textit{FDR} than CodeLlama-7b/Cosine (mean~=~$0.80$, median~=~$0.81$) and CodeLlama-7b/Euclidean (mean~=~$0.76$, median~=~$0.76$). This suggests that language models with fewer parameters ($125 Million$ for UniXcoder) can outperform much larger language models ($7 Billion$ for CodeLlama-7b) for our minimization problem. This can be attributed to the fact that CodeLlama-7b, like most large language models, is a decoder-only model that was pre-trained for code generation tasks, while UniXcoder employs code understanding tasks that are more effective in analyzing and comprehending the contextual semantic information from the source code~\cite{fan2023large}.

    Recall that the GA terminates when the fitness value improves by less than $0.0025$, thus resulting in different numbers of generations for each configuration across project versions. The overall average number of generations of all configurations across versions is $47.73$, with UniXcoder/Cosine having a higher average ($63.10$ generations). However, if a fixed number of generations had been used, say $40$ generations, UniXcoder/Cosine would still yield the highest \textit{FDR} ($0.81$) among all configurations. This clearly makes UniXcoder/Cosine the best \ourapproach~alternative, in terms of \textit{FDR}. 
    
\begin{table*}
   \caption{Results and descriptive statistics of \textit{FDR} of \ourapproach~and ATM across projects for the $50\%$ minimization budget. The highest \textit{FDR} results are highlighted in bold.}
   \vspace{-10pt}
    \newcommand{\cellwithlinebreak}[2][c]{\begin{tabular}[#1]{@{}c@{}}#2\end{tabular}}
    \renewcommand{\arraystretch}{1.1}
    \centering
     \resizebox{1\textwidth}{!}{
     \begin{tabular}{l|c|c|c|c|c|c|c|c|c|c|c|c|c}
     \hline 
     \multirow{2}{*}{\diagbox[dir=SE, height=2.2\line]{\raisebox{1ex}{~\textbf{Project}}}{\raisebox{-0.2ex}{\textbf{Approach}}}} &
     \multicolumn{2}{c}{CodeBERT} & \multicolumn{2}{|c}{GraphCodeBERT} & \multicolumn{2}{|c}{UniXcoder} & \multicolumn{2}{|c}{StarEncoder} & \multicolumn{2}{|c|}{CodeLlama-7b} & UniXcoder-p & ATM & ATM \\
      \cline{2-14}
      & Cosine  & Euclidean  & Cosine  & Euclidean  & Cosine  & Euclidean & Cosine  & Euclidean & Cosine  & Euclidean & Cosine & Combined  & Tree Edit Distance  \\\cline{2-14}
       \hline
       Chart           & 0.82 & 0.84 & 0.79 & 0.75 & 0.82 & 0.69 & 0.71 & 0.79 & 0.72 & 0.73 & 0.77 & 0.88 & \textbf{0.89} \\
       Cli             & 0.81 & 0.81 & 0.83 & 0.86 & 0.82 & 0.81 & 0.83 & 0.86 & \textbf{0.87} & 0.86 & 0.80 & \textbf{0.87} & \textbf{0.87} \\
       Codec           & 0.89 & \textbf{0.90} & 0.83 & 0.88 & 0.80 & 0.81 & 0.82 & 0.83 & 0.87 & 0.78 &0.78 & 0.77 & 0.80 \\
       Collections     & 0.70 & 0.82 & 0.95 & 0.75 & 0.95 & 0.70 & 0.70 & 0.75 & 0.78 & 0.70 & \textbf{1.00} & 0.98 & 0.92 \\
       Compress        & 0.85 & 0.86 & 0.87 & 0.87 & 0.86 & 0.81 & 0.78 & 0.84 & 0.85 & 0.79 & \textbf{0.89} & 0.88 & \textbf{0.89} \\
       Csv             & 0.86 & 0.87 & 0.88 & 0.91 & 0.85 & 0.79 & 0.84 & 0.86 & \textbf{0.93} & 0.88 & 0.88 & 0.92 & 0.91 \\
       Gson            & 0.70 & 0.71 & 0.72 & 0.75 & \textbf{0.81} & 0.75 & 0.76 & 0.76 & \textbf{0.81} & 0.77 &0.76 & 0.76 & 0.77 \\
       JacksonCore     & 0.70 & 0.73 & 0.73 & 0.73 & 0.75 & 0.67 & 0.74 & 0.74 & 0.75 & 0.75 & \textbf{0.78} & 0.69 & 0.72 \\
       JacksonDatabind & 0.66 & 0.69 & 0.68 & 0.67 & \textbf{0.76} & 0.65 & 0.65 & 0.69 & 0.67 & 0.65 &0.74 & 0.69 & 0.70 \\
       JacksonXml      & 0.77 & 0.85 & 0.83 & 0.73 & 0.83 & 0.82 & 0.67 & 0.65 & 0.83 & 0.85 & \textbf{0.87} & 0.60 & 0.73 \\
       Jsoup           & 0.75 & 0.74 & 0.77 & 0.76 & \textbf{0.80}  & 0.76 & 0.73 & 0.75 & 0.77 & 0.75 & 0.76& 0.70 & 0.69 \\
       JxPath          & 0.88 & 0.90 & 0.92 & 0.86 & \textbf{0.94} & 0.83 & 0.72 & 0.77 & 0.78 & 0.75 & \textbf{0.94} & 0.79 & 0.79 \\
       Lang            & 0.81 & 0.85 & 0.86 & 0.75 & \textbf{0.90}  & 0.71 & 0.77 & 0.85 & 0.83 & 0.71 &0.84 & 0.78 & 0.83 \\
       Math            & 0.74 & 0.78 & 0.75 & 0.74 & 0.78 & 0.65 & 0.71 & 0.78 & 0.75 & 0.67 & 0.77& 0.81 & \textbf{0.82} \\
       Mockito         & 0.80 & 0.83 & 0.84 & 0.81 & \textbf{0.91} & 0.79 & 0.77 & 0.83 & 0.80 & 0.77 &0.78 & 0.77 & 0.79 \\
       Time            & 0.83 & 0.88 & 0.85 & 0.82 & 0.82 & 0.77 & 0.82 & 0.79 & 0.81 & 0.76 &0.86 & \textbf{0.89} & 0.88 \\
        &&&&&&&&&&&&&\\[-8pt]
       \hline
       \textbf{~~~~~~~~~~~Statistics}&&&&&&&&&&&&\\[-3pt]
Min            & 0.66     & 0.69    & 0.68    & 0.67   & \textbf{0.75}   & 0.65    & 0.65    & 0.65    & 0.67   & 0.65  &0.74 & 0.60   & 0.69   \\
25\% Quantile  & 0.73     & 0.77    & 0.77    & 0.75   & \textbf{0.80}   & 0.70    & 0.71    & 0.75    & 0.77   & 0.73  &0.77& 0.75   & 0.76   \\
Mean           & 0.79     & 0.82    & 0.82    & 0.79   & \textbf{0.84}   & 0.75    & 0.75    & 0.78    & 0.80   & 0.76 & 0.83 & 0.80   & 0.81   \\
Median         & 0.81     & \textbf{0.84}    & 0.83    & 0.76   & 0.82   & 0.77    & 0.75    & 0.79    & 0.81   & 0.76 & 0.79 & 0.79   & 0.81   \\
75\% Quantile  & 0.84     & 0.86    & 0.86    & 0.86   & 0.87   & 0.81    & 0.79    & 0.83    & 0.84   & 0.78 & 0.87 & \textbf{0.88}   & \textbf{0.88}   \\
Max            & 0.89     & 0.90    & 0.95    & 0.91   & 0.95   & 0.83    & 0.84    & 0.86    & 0.93   & 0.88 & \textbf{1.00} & 0.98   & 0.92   \\
        \hline
     \end{tabular}
     }
     \label{tab:FDR_results}
     \vspace{-3pt}
\end{table*}

\begin{table*}
   \caption{Results and descriptive statistics of \textit{MT} (in min) of \ourapproach~and ATM across projects for the $50\%$ minimization budget. The shortest \textit{MT} results are highlighted in bold.}
   \vspace{-10pt}
    \newcommand{\cellwithlinebreak}[2][c]{\begin{tabular}[#1]{@{}c@{}}#2\end{tabular}}
    \renewcommand{\arraystretch}{1.1}
    \centering
     \resizebox{1\textwidth}{!}{
     \begin{tabular}{l|c|c|c|c|c|c|c|c|c|c|c|c|c}
     \hline 
     \multirow{2}{*}{\diagbox[dir=SE, height=2.2\line]{\raisebox{1ex}{~\textbf{Project}}}{\raisebox{-0.2ex}{\textbf{Approach}}}} &
     \multicolumn{2}{c}{CodeBERT} & \multicolumn{2}{|c}{GraphCodeBERT} & \multicolumn{2}{|c}{UniXcoder} & \multicolumn{2}{|c}{StarEncoder} & \multicolumn{2}{|c|}{CodeLlama-7b} & UniXcoder-p & ATM~~~~ & ATM~~~~~~~~~~~\\
      \cline{2-14}
      & Cosine  & Euclidean  & Cosine  & Euclidean  & Cosine  & Euclidean & Cosine  & Euclidean & Cosine  & Euclidean & Cosine & Combined  & Tree Edit Distance  \\\cline{2-14}
       \hline
Chart           & \textbf{0.92} & 1.51 & 0.97 & 1.20  & 1.33 & 0.99 & 2.24 & 2.68  & 21.02  & 20.92  & 2.18 & 7.86  & 94.22  \\
Cli             & \textbf{0.11} & 0.16 & 0.14 & 0.15 & 0.17 & 0.14 & 0.16 & 0.22  & 4.39  & 4.38  & 0.22 & 0.23  & 1.20   \\
Codec           & \textbf{0.15} & 0.21 & 0.18 & 0.17 & 0.22 & 0.18 & 0.25 & 0.33  & 5.75  & 5.65 & 0.31 & 0.42  & 3.02   \\
Collections     & \textbf{0.37} & 0.51 & 0.43 & 0.50  & 0.58 & 0.38 & 0.79 & 1.11  & 8.96  & 8.80  & 0.87 & 2.16  & 14.89  \\
Compress        & \textbf{0.16} & 0.20  & 0.19 & 0.18 & 0.22 & 0.17 & 0.25 & 0.34  & 9.26  & 9.18  & 0.32 & 0.55  & 6.55   \\
Csv             & \textbf{0.10} & 0.13 & 0.12 & 0.11 & 0.15 & 0.11 & 0.14 & 0.16  & 4.19  & 4.15  & 0.19 & 0.16  & 0.35   \\
Gson            & \textbf{0.31} & 0.39 & \textbf{0.31} & 0.33 & 0.47 & 0.32 & 0.68 & 0.79  & 8.31  & 8.26  & 0.77 & 1.48  & 6.09   \\
JacksonCore     & \textbf{0.14} & 0.16 & 0.16 & 0.15 & 0.19 & 0.15 & 0.22 & 0.25  & 11.10  & 11.06 & 0.27 & 0.35  & 2.13   \\
JacksonDatabind & 0.92 & 1.40  & \textbf{0.91} & 1.35 & 1.43 & 1.33 & 2.15 & 2.46  & 34.67 & 35.41 & 2.50 & 4.90  & 22.69  \\
JacksonXml      & 0.11 & 0.11 & 0.11 & \textbf{0.10} & 0.13 & 0.11 & 0.12 & 0.13  & 6.86  & 6.84 & 0.18 & 0.13  & 0.34   \\
Jsoup           & \textbf{0.17} & \textbf{0.17} & \textbf{0.17} & \textbf{0.17} & 0.22 & 0.18 & 0.29 & 0.29  & 20.92  & 20.91  & 0.33 & 0.55  & 1.63   \\
JxPath          & \textbf{0.11} & 0.15 & 0.13 & 0.13 & 0.17 & 0.13 & 0.17 & 0.21  & 4.98  & 4.95  & 0.23 & 0.22  & 0.73   \\
Lang            & \textbf{0.91} & 1.47 & \textbf{0.91} & 1.08 & 1.38 & 0.97 & 2.35 & 2.82  & 15.09  & 14.98  & 2.53 & 5.44  & 33.47  \\
Math            & 1.33 & 2.33 & \textbf{1.31} & 2.06 & 2.12 & 1.98 & 3.41 & 4.22  & 21.12 & 23.09 & 3.69 & 12.05 & 215.43 \\
Mockito         & \textbf{0.40} & 0.50 & 0.42 & \textbf{0.40} & 0.59 & 0.41 & 0.97 & 1.12  & 17.40  & 17.93  & 1.15 & 2.32  & 9.34   \\
Time            & 3.49 & 5.35 & \textbf{3.43} & 3.91 & 3.73 & 3.44 & 9.06 & 10.45 & 41.75  & 39.60 & 7.61 & 26.13 & 163.18 \\
        &&&&&&&&&&&&&\\[-8pt]
       \hline
       \textbf{~~~~~~~~~~~Statistics}&&&&&&&&&&&&\\[-3pt]
Min    & \textbf{0.10}    & 0.11     & 0.11     & \textbf{0.10}      & 0.13    & 0.11     & 0.12     & 0.13    & 4.19     & 4.15    & 0.18 & 0.13     & 0.34     \\
25\% Quantile   & \textbf{0.13}  & 0.16     & 0.16    & 0.15     & 0.19   & 0.15   & 0.21   & 0.24  & 6.58   & 6.54    & 0.26 & 0.32     & 1.52   \\
Mean   & \textbf{0.61} & 0.92 & 0.68 & 0.75 & 0.82 & 0.69 & 1.45 & 1.72 & 14.74 & 14.72 & 1.46 & 4.06 & 35.95 \\
Median & \textbf{0.24}    & 0.30      & 0.25     & 0.26    & 0.35   & 0.25     & 0.49    & 0.57   & 10.18    & 10.12    & 0.55 & 1.02    & 6.32     \\
75\% Quantile   & \textbf{0.91}  & 1.42   & \textbf{0.91}     & 1.11     & 1.34  & 0.98    & 2.17   & 2.52   & 20.95     & 20.91 & 2.26 & 5.04    & 25.39   \\
Max    & 3.49    & 5.35     & \textbf{3.43}     & 3.91     & 3.73    & 3.44     & 9.06     & 10.45   & 41.75     & 39.60   & 7.61 & 26.13    & 215.43   \\
        \hline
     \end{tabular}
     }
     \label{tab:MT_results}
     \vspace{-3pt}
\end{table*}

\begin{table*}
   \caption{Results and descriptive statistics of \textit{TSR} (in percentage) of \ourapproach~and ATM across projects for the $50\%$ minimization budget. The highest \textit{TSR} results are highlighted in bold.}
   \vspace{-10pt}
        \newcommand{\cellwithlinebreak}[2][c]{\begin{tabular}[#1]{@{}c@{}}#2\end{tabular}}
    \renewcommand{\arraystretch}{1.1}
    \centering
     \resizebox{1\textwidth}{!}{
     \begin{tabular}{l|c|c|c|c|c|c|c|c|c|c|c|c|c}
     \hline 
     \multirow{2}{*}{\diagbox[dir=SE, height=2.2\line]{\raisebox{1ex}{~\textbf{Project}}}{\raisebox{-0.2ex}{\textbf{Approach}}}} &
     \multicolumn{2}{c}{CodeBERT} & \multicolumn{2}{|c}{GraphCodeBERT} & \multicolumn{2}{|c}{UniXcoder} & \multicolumn{2}{|c}{StarEncoder} & \multicolumn{2}{|c|}{CodeLlama-7b} & UniXcoder-p & ATM & ATM\\
      \cline{2-14}
      & Cosine  & Euclidean  & Cosine  & Euclidean  & Cosine  & Euclidean & Cosine  & Euclidean & Cosine  & Euclidean & Cosine & Combined  & Tree Edit Distance  \\\cline{2-14}
       \hline
Chart           & 55.45\% & 57.35\% & 58.78\% & 59.81\% & 58.03\% & 56.11\% & 58.47\% & \textbf{63.08\%} & 56.64\% & 59.14\% & 55.27\% & 59.77\% & 57.53\% \\
Cli             & 29.90\%  & 27.90\%  & \textbf{30.54\%} & 29.75\% & 28.32\% & 28.17\% & 29.52\% & 27.52\% & 28.25\% & 28.46\% & 29.95\%& 28.07\% & 27.34\% \\
Codec           & 44.11\% & 40.68\% & 47.59\% & 49.93\% & \textbf{55.58\%} & 50.62\% & 51.05\% & 48.38\% & 45.27\% & 48.83\% & 45.84\% & 34.62\% & 45.94\% \\
Collections     & 57.40\%  & 55.05\% & 56.21\% & 56.10\%  & 55.41\% & 59.05\% & \textbf{59.41\%} & 57.36\% & 55.01\% & 56.78\% & 56.37\% & 53.57\% & 57.65\% \\
Compress        & 37.74\% & 37.98\% & 36.56\% & 38.10\%  & 39.10\%  & \textbf{40.70\%}  & 37.87\% & 35.26\% & 36.45\% & 40.11\% & 36.33\% & 33.85\% & 32.59\% \\
Csv             & 18.66\% & 18.78\% & 19.90\%  & 20.07\% & 18.13\% & 10.11\% & 16.40\%  & 16.15\% & 12.82\% & 9.37\%  & 6.92\%& \textbf{20.45\%} & 19.89\% \\
Gson            & 41.81\% & 40.60\%  & 42.16\% & 42.01\% & 41.40\%  & \textbf{44.63\%} & 41.85\% & 40.10\%  & 41.69\% & 43.63\% & 39.51\%& 40.35\% & 38.17\% \\
JacksonCore     & 47.45\% & \textbf{48.44\%} & 44.37\% & 46.18\% & 47.10\%  & 48.00\%  & 46.02\% & 46.42\% & 46.95\% & 46.98\% & 45.22\%& 43.59\% & 46.71\% \\
JacksonDatabind & 42.54\% & 41.28\% & 41.68\% & 41.63\% & 38.32\% & 42.37\% & 41.71\% & 39.56\% & 41.29\% & 41.75\% & 39.98\%& \textbf{50.61\%} & 47.92\% \\
JacksonXml      & 42.81\% & 42.05\% & 42.63\% & 44.12\% & 45.80\%  & 46.32\% & 45.70\%  & \textbf{49.52\%} & 44.03\% & 43.60\%  & 44.48\%& 39.17\% & 39.34\% \\
Jsoup           & 42.96\% & 42.03\% & 41.33\% & \textbf{43.80\%}  & 33.90\%  & 41.65\% & 38.60\%  & 35.61\% & 35.12\% & 37.55\% & 32.55\%& 40.39\% & 38.20\%  \\
JxPath          & 43.37\% & 42.82\% & 41.00\%  & 41.90\%  & 40.83\% & 42.70\%  & 42.46\% & 41.84\% & 44.12\% & 44.16\% &41.04\% & 44.40\%  & \textbf{44.46\%} \\
Lang            & 38.36\% & 36.49\% & 40.14\% & 41.24\% & 34.05\% & 44.76\% & 41.04\% & 35.37\% & 42.83\% & \textbf{44.97\%} & 32.19\%& 37.71\% & 38.06\% \\
Math            & 50.24\% & 49.43\% & 50.26\% & 49.47\% & 49.67\% & 49.67\% & 50.00\%  & 49.72\% & 50.90\%  & \textbf{51.34\%} & 45.44\% & 47.93\% & 49.10\%  \\
Mockito         & 27.65\% & 29.87\% & 34.33\% & 38.38\% & 33.97\% & 39.93\% & \textbf{44.82\%} & 41.80\%  & 32.13\% & 39.77\% & 24.12\% & 24.31\% & 26.13\% \\
Time            & 48.12\% & 47.57\% & 48.33\% & 48.49\% & 47.89\% & \textbf{49.06\%} & 47.35\% & 46.94\% & 46.84\% & 48.38\% & 46.17\%& 45.93\% & 47.35\% \\

        &&&&&&&&&&&&&\\[-8pt]
       \hline
       \textbf{~~~~~~~~~~~Statistics}&&&&&&&&&&&&\\[-3pt]
Min  & 18.66\%   & 18.78\%  & 19.90\%   & 20.07\%   & 18.13\%   & 10.11\%   & 16.40\%    & 16.15\%   & 12.82\%   & 9.37\%   & 6.92\%& \textbf{20.45\%}   & 19.89\%   \\
25\% Quantile & 38.21\%  & 37.61\% & 39.25\%  & 40.53\%  & 34.03\%   & \textbf{41.41\%} & 40.43\%   & 35.55\%   & 36.12\% & 40.03\%  & 32.46\%& 34.43\% & 36.69\% \\
Mean & 41.79\% & 41.15\%  & 42.24\% & 43.19\% & 41.72\% & \textbf{43.37\%} & 43.27\% & 42.16\% & 41.27\% & 42.80\% & 38.84\% & 40.30\%  & 41.02\% \\
Median & 42.89\%  & 41.66\%  & 41.92\%   & 42.91\%  & 41.12\%  & \textbf{44.70\%}  & 43.64\%   & 41.82\%   & 43.43\%   & 43.90\%  & 40.51\% & 40.37\%   & 41.90\%    \\
75\% Quantile & 47.62\% & 47.79\% & 47.78\%  & 48.74\%  & 48.34\%  & \textbf{49.21\%} & 48.01\% & 48.67\%  & 46.87\% & 48.49\% & 45.54\% & 46.43\%   & 47.49\% \\
Max  & 57.40\%    & 57.35\%   & 58.78\%   & 59.81\%   & 58.03\%   & 59.05\%   & 59.41\%   & \textbf{63.08\%}   & 56.64\%   & 59.14\%   & 56.37\% & 59.77\%   & 57.65\%   \\
        \hline
     \end{tabular}
     }
     \label{tab:TSR_results}
     \vspace{-3pt}
\end{table*}

\begin{table*}
   \caption{Results of \textit{FDR} (in percentage), \textit{MT} (in min) and \textit{TSR} (in percentage) of \ourapproach~for the $Closure$ project and a $50\%$ minimization budget. The highest \textit{FDR} and \textit{TSR} as well as the shortest \textit{MT} are highlighted in bold.}
   \vspace{-10pt}
        \newcommand{\cellwithlinebreak}[2][c]{\begin{tabular}[#1]{@{}c@{}}#2\end{tabular}}
    \renewcommand{\arraystretch}{1.1}
    \centering
     \resizebox{1\textwidth}{!}{
     \begin{tabular}{l|c|c|c|c|c|c|c|c|c|c}
     \hline 
     \multirow{2}{*}{\diagbox[dir=SE, height=2.2\line]{\raisebox{1ex}{~\textbf{Metric}}}{\raisebox{-0.2ex}{\textbf{Approach}}}} &
     \multicolumn{2}{c}{CodeBERT} & \multicolumn{2}{|c}{GraphCodeBERT} & \multicolumn{2}{|c}{UniXcoder} & \multicolumn{2}{|c}{StarEncoder} & \multicolumn{2}{|c}{CodeLlama-7b} \\
      \cline{2-11}
      & Cosine  & Euclidean  & Cosine  & Euclidean  & Cosine  & Euclidean & Cosine  & Euclidean & Cosine  & Euclidean   \\\cline{2-11}
       \hline
        FDR & 0.77 & 0.77 & 0.78 & 0.78 & \textbf{0.79} & 0.74 & 0.73 & 0.76 & 0.77 & 0.76\\
        MT & \textbf{15.02} & 30.53 & 19.73 & 25.98 & 17.80 & 27.61 & 16.45 & 34.17 &  54.87 & 66.84\\
        TSR & 53.18\% & 51.98\% & 53.75\% & 52.53\% & 52.55\% & 54.27\% & \textbf{54.40\%}  &  52.51\%& 53.81\% & 53.70\% \\
       \hline
     \end{tabular}
     }
     \label{tab:Closure_results}
\end{table*}

\textbf{\textit{Achieved Saving in Testing Time.}}
Table~\ref{tab:TSR_results} shows that, for the $50\%$ minimization budget, {\ourapproach} achieved an average \textit{TSR} ranging from $41.15\%$ to $43.37\%$, across configurations. The average \textit{TSR} achieved by UniXcoder/Cosine was $41.72\%$. This indicates that, reducing the number of test cases by $50\%$ results in a $41.72\%$ reduction in testing time.

\vspace{4pt}
\noindent\textbf{\textit{RQ1.2 results.}}
Table~\ref{tab:MT_results} shows that the average \textit{MT} of all \ourapproach~configurations ranges from $0.61$ to $14.74$ min per project version.
The \textit{MT} of \ourapproach~using UniXcoder/Cosine, which achieved the highest average \textit{FDR} ($0.84$), also appears to be efficient with $mean~=~0.82~min$ and $median~=~0.35~min$.

\vspace{3pt}
\textbf{\textit{Preparation Time.}}
The average preparation time ranges from $0.26$ to $17.71$ min per project version, across configurations.
We observed that the preparation time of UniXcoder/Cosine is $65.43$ times shorter than CodeLlama-7b. This is because (1) larger language models require much longer model loading time ($12.07$ min for CodeLlama-7b, compared to UniXcoder for $0.04$ min, on average) and significantly larger memory ($13.10GB$ for CodeLlama-7b compared to $504MB$ for UniXcoder) and (2) the embeddings generation time of larger language models is significantly longer ($5.15$ min for CodeLlama-7b compared to $0.16$ min for UniXcoder, on average) as they have more complex architectures and a larger number of parameters, making the generation of embeddings more time-consuming.

\vspace{3pt}
\textbf{\textit{Search Time.}}
The average search time of {\ourapproach} ranges from $0.46$ to $1.90$ min per project version, across configurations. \ourapproach~using CodeBERT with Cosine similarity was the fastest in terms of search time, ranging from $0.04$ to $3.86$ min, across project versions. Not only \ourapproach~using UniXcoder/Cosine took a shorter search time ($0.78$ min) than the average search time across configurations ($1.03$ min), it also achieved the highest \textit{FDR} among all configurations, thus making it the best option for \ourapproach.

\textbf{\textit{Results for the Closure project.}}
We further evaluated \ourapproach~on the $Closure$ project, which has $174$ versions with an average of $7,308$ test cases per version. Though this project is part of the \textsc{Defects4J} dataset, it was previously excluded from the ATM evaluation due to its scalability limitations~\cite{pan2023atm} and the large size of our experiments. However,  evaluating the effectiveness and efficiency of \ourapproach~on this project, considering its larger scale and test suite, can be informative. Table~\ref{tab:Closure_results} presents the results regarding \textit{FDR}, \textit{MT} (in min), and \textit{TSR} (in percentage), respectively. We can see that UniXcoder/Cosine achieves once again the highest \textit{FDR} of $0.79$ and an \textit{MT} of only $17.80~min$, while saving $52.55\%$ testing time, which indicates that \ourapproach~can be effective and scalable for larger projects and test suites.

\summarybox{\textbf{RQ1 summary.}
    \ourapproach~achieved high \textit{FDR} results ($0.79$, on average, across configurations) for a $50\%$ minimization budget. UniXcoder/Cosine is the best \ourapproach~configuration when considering both effectiveness ($0.84$ \textit{FDR} on average) and efficiency ($0.82$ min on average), with an average \textit{TSR} of $41.72\%$.}

\subsubsection{RQ2 results}
\label{RQ2_results}
~

\noindent\textbf{\textit{RQ2.1 results.}}
For a $50\%$ minimization budget, Tables~\ref{tab:FDR_results}, \ref{tab:MT_results}, and \ref{tab:TSR_results} report  \textit{FDR}, \textit{MT}, and \textit{TSR}, respectively, for the best configuration of \ourapproach~(i.e., UniXcoder/Cosine) and the two best configurations of the baseline approach: ATM with combined similarity and tree edit distance similarity. 

We observe that \ourapproach~significantly outperforms ATM/Combined, the best ATM alternative, and ATM/TreeEditDistance in terms of \textit{FDR}, with $\alpha = 0.01$. 
Compared to ATM, \ourapproach~achieves higher average \textit{FDR} ($0.84$) with lower standard deviation ($0.06$ compared to $0.10$ for ATM/Combined and $0.08$ for ATM/TreeEditDistance).

Figure~\ref{fig:convergence_plot} shows that UniXcoder/Cosine converges faster to a higher \textit{FDR} ($0.84$ across projects), compared to ATM/Combined ($0.80$ across projects) and ATM/TreeEditDistance ($0.81$ across projects). 
This may be explained by the fact that (1) UniXcoder can handle full code (i.e., the source code of test cases without preprocessing), whereas ATM uses preprocessed code; and (2) UniXcoder was pre-trained on various code understanding tasks, enabling the model to learn contextual information for each token and the meaning of each code element (e.g., method, variable names, and string characters) in the input test code. In contrast, ATM relies on tree-based similarity based on ASTs, which lack information regarding the syntactic context and certain details of the test code.

For some projects, ATM achieved higher \textit{FDR} than \ourapproach. However, a Fisher's Exact Test suggests that, for these projects, there is no significant difference between the \textit{FDR} results of ATM and \ourapproach. Nevertheless, a possible explanation for such performance difference could be that the percentage of test code exceeding the token limits of language models on these projects (5.72\%, 3.21\%, 3.25\%, and 7.25\% for $Chart$, $Compress$, $Collections$, and $Math$, respectively) is higher than the average of 2.98\% across projects, and that the truncated parts are important for distinguishing test cases. Another possible explanation could be due to variability across test cases in some projects (e.g., $Cli$ and $Time$), regarding their low-level code elements such as complex string manipulation and domain-specific character strings, rather than structural information. This makes tree-based similarity better at capturing these differences, by comparing AST node labels, than using the Cosine Similarity of embeddings.

Note that the way ATM preprocessed test cases, by removing all the comments and assertions, may also influence its performance. For example, for the $Codec$ project, 26.18\% of the test methods contain comments, which is higher than the average percentage across projects (14.85\%). Such comments may convey rich information about test code, such as test case descriptions and expected results. 
Therefore, due to such loss of information in code comments, ATM may not efficiently achieve higher \textit{FDR} for this project.

The average number of generations required by \ourapproach~was $63.10$ across project versions (maximum~=~$100$), which is lower than ATM/Combined (mean~=~$85.60$ and maximum~=~$130$) and ATM/TreeEditDistance (mean~=~$74.88$ and maximum~=~$115$). This indicates that \ourapproach~converges faster across generations than ATM, in terms of fitness value, which is the sum of the maximum squared similarity values for all test case pairs per version. 
Related to this, if due to time constraints a fixed number of generations is used as the termination criterion for search, say $40$ generations, it is clear based on Figure~\ref{fig:convergence_plot} that UniXcoder/Cosine would yield the highest \textit{FDR}. This indicates that \ourapproach~offers better guidance for the GA search, making it more effective at removing redundant test cases while maintaining a higher fault detection rate, thus explaining the results in Table~\ref{tab:FDR_results} and why \ourapproach~is a better option than ATM for test suite minimization in practice. 

\textbf{\textit{Achieved Saving in Testing Time.}}
The average time saving rate of the best configuration of {\ourapproach} ranges from $18.13\%$ to $58.03\%$ across projects, with a mean of $41.72\%$ across projects. We observe that this time saving rate is slightly higher than, but not significantly different from, that of ATM/Combined ($40.30\%$) and ATM/TreeEditDistance ($41.02\%$), as shown by the Wilcoxon signed-rank test.

\textbf{\textit{Comparing \ourapproach~and ATM using preprocessed code.}}
As discussed in Section~\ref{sec:baseline}, ATM cannot handle full code due to scalability issues. However, the preprocessing step might remove useful information from the test code and thus impact the effectiveness of ATM. Therefore, in order to explain the {\ourapproach} \textit{FDR} results when compared to ATM, we want to determine how the use of full code and the {\ourapproach} test method embeddings individually affect \textit{FDR}.
For this, we conducted an additional experiment to assess {\ourapproach} with UniXcoder/Cosine, the best {\ourapproach} configuration, using preprocessed code, on all $16$ projects.
Tables~\ref{tab:FDR_results} and \ref{tab:TSR_results} report the \textit{FDR} and \textit{TSR} results, respectively, for this configuration using preprocessed code (UniXcoder-p) and the two best configurations of ATM (using GA with combined and tree edit distance similarity measures), for a $50\%$ minimization budget. 
We observe that, using the same input as ATM (preprocessed test code), \ourapproach~with UniXcoder/Cosine still yields a higher average \textit{FDR} ($0.83$) than ATM ($0.81$). Still, it is lower than what we achieved using {\ourapproach} with full code ($0.84$).
The average \textit{TSR} result for {\ourapproach} with preprocessed code is however slightly lower than that of ATM ($38.84\%$ versus $41.02\%$), though the difference is not statistically significant, as resulted by the Wilcoxon signed-rank test.
These results therefore suggest that the small improvements in \textit{FDR} observed with {\ourapproach} are due to both {\ourapproach}'s test method embeddings and its capacity to handle full code in a scalable way. \textit{TSR} remains similar for {\ourapproach} and ATM, with or without preprocessing. We report the results for {\ourapproach} with UniXcoder/Cosine using preprocessed code on all $16$ projects in our replication package~\cite{our_replication_package}. 

\vspace{5pt}
\noindent\textbf{\textit{RQ2.2 results.}}

\begin{figure*}[ht]
    \centering
    \includegraphics[width=0.90\textwidth]{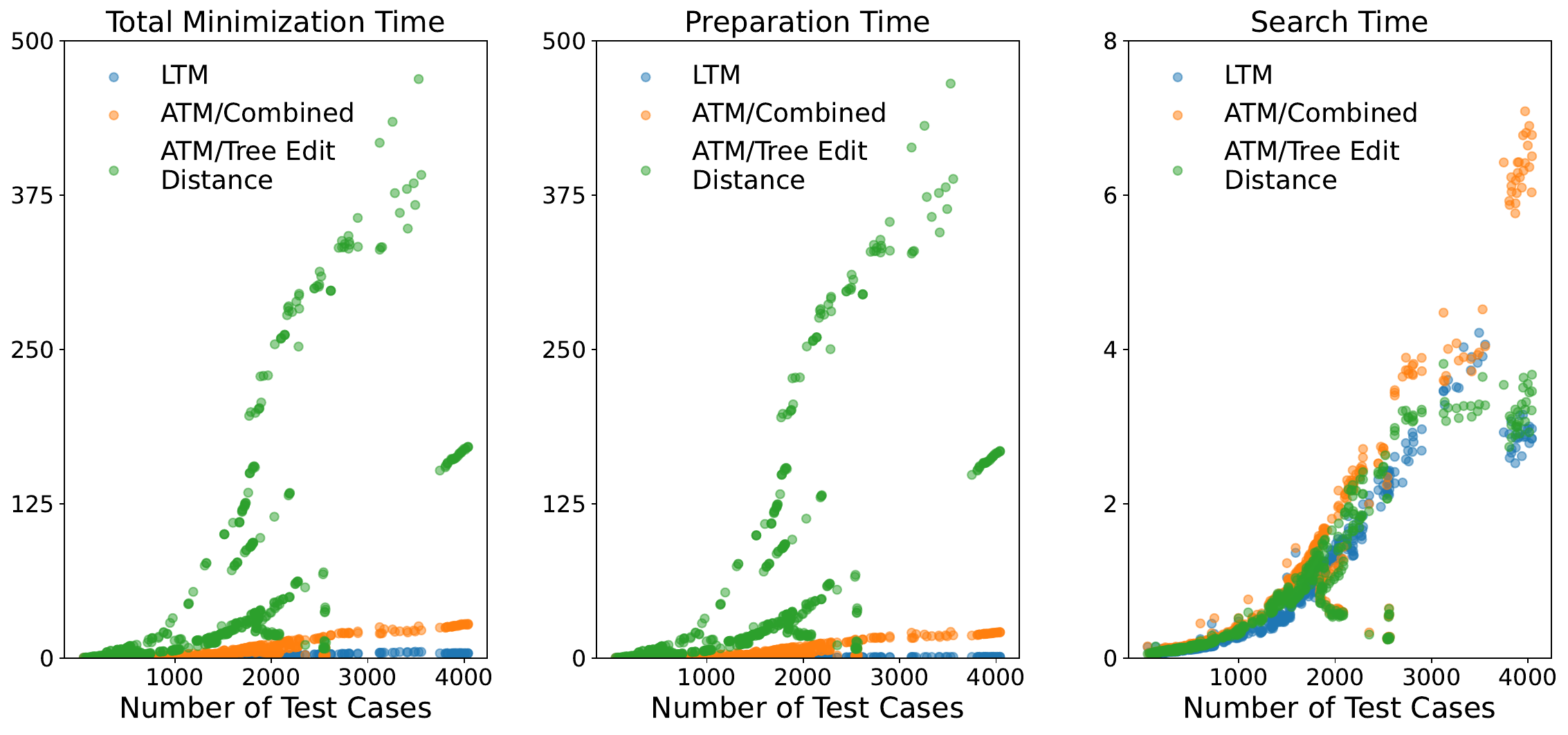}
    \vspace{-5pt}
    \caption{Scatter plots of the number of test cases and \textit{MT}, preparation time, and search time (in min), for \ourapproach~(UniXcoder/Cosine) and ATM, across all the $661$ project versions for the 50\% minimization budget}
    \vspace{-8pt}
    \label{fig:scatter_time_rq2}
\end{figure*}

Figure~\ref{fig:scatter_time_rq2} depicts \textit{MT}, preparation time, and search time---the first one including the last two---as a function of the number of test cases per version, for the $661$ versions of the $16$ projects. This is done for the best configuration of \ourapproach~(i.e., UniXcoder/Cosine) and compared to the best configurations of ATM (ATM/Combined and ATM/TreeEditDistance) for the $50\%$ minimization budget.

We observe that \ourapproach~ran much faster than ATM in terms of \textit{MT} ($79.80\%$ and $97.72\%$ faster than ATM/Combined and ATM/TreeEditDistance on average, respectively) with both much less preparation time ($92.70\%$ and $99.35\%$ less than ATM/Combined and ATM/TreeEditDistance on average, respectively) and search time ($35.16\%$ and $7.81\%$ less than ATM/Combined and ATM/TreeEditDistance on average, respectively). 
In addition, we observe that the observations in Figure~\ref{fig:scatter_time_rq2} follow a quadratic relationship rather than a linear one. Therefore, a quadratic regression model is fitted for the preparation time, search time, and \textit{MT} as a function of the number of test cases per version, respectively, to quantify and compare relationships across approaches.

The equation of the quadratic regression model is as follows.\vspace{-10pt}

    \begin{equation}
    Time = a * n^2 + b * n + c
    \end{equation}
where $n$ denotes the number of test cases per project version, $a$ and $b$ are regression coefficients, and $c$ denotes the intercept of the quadratic regression model.

The results of quadratic regression confirmed that the \textit{MT} and the number of test cases, for both \ourapproach~and ATM, follow a quadratic relationship as the coefficients on the quadratic term were statistically significant with $\alpha = 0.01$. This is, as expected, due to the fact that (1) similarity values are measured between test case pairs, which increases quadratically with the number of test cases, and (2) the search space for GA increases rapidly with the number of test cases. 
The $R^2$ values for the regression model of \ourapproach~and ATM/Combined were $98.89\%$ and $89.51\%$, respectively. This indicates that a high percentage of the variation in \textit{MT} can be explained by the number of test cases. However, the $R^2$ value for ATM/TreeEditDistance was only $50.36\%$, indicating that a much smaller proportion of the variation in \textit{MT} can be explained by the number of test cases in this case. This is because the preparation time of ATM/TreeEditDistance, which is part of its \textit{MT}, was also largely influenced by AST sizes. More discussions will be provided in the following section.

Moreover, a comparison of regression coefficients indicates that the \textit{MT} of \ourapproach~increases up to two orders of magnitude slower than ATM with the squared number of test cases per project version: $8.255e^{-06}$ for \ourapproach~versus $9.314e^{-05}$ and $4.446e^{-04}$ for ATM/Combined and ATM/TreeEditDistance, respectively. For large industry projects with large test suites, \ourapproach~is thus much more scalable in terms of \textit{MT}, and therefore a better option than ATM in practice.

\vspace{2pt}
\textbf{\textit{Preparation Time.}}
We found that \ourapproach~is much more efficient, in terms of preparation time across project versions (mean~=~$0.23$ min and median~=~$0.14$ min), compared to ATM/Combined (mean~=~$3.15$ min and median~=~$0.74$ min) and ATM/TreeEditDistance (mean~=~$35.31$ min and median~=~$6.10$ min). Moreover, the preparation time of \ourapproach~takes only $35.75\%$ of the \textit{MT} per version, which is much less than ATM/Combined ($71.22\%$) and ATM/TreeEditDistance ($93.74\%$).
Quadratic regression shows that the coefficient on the quadratic term for \ourapproach~($1.241e^{-06}$) is one order of magnitude smaller than the coefficients for ATM/Combined ($6.634e^{-05}$) and two orders of magnitude smaller than ATM/TreeEditDistance ($4.3680e^{-04}$), with all coefficients statistically significant with $\alpha = 0.01$. This indicates that, on larger industry projects with much larger test suites, the absolute calculation time difference between \ourapproach~and ATM is expected to be extremely large, thus making \ourapproach~a better option regarding scalability in preparation time.

The $R^2$ values for the regression model of \ourapproach~and ATM/Combined are $97.70\%$ and $87.86\%$, respectively. This indicates that more than $87\%$ of the variation in preparation time can be explained by the number of test cases. However, the $R^2$ value for ATM/TreeEditDistance is 
$49.96\%$, which is mainly due to its preparation time as it is highly influenced by test code length. For instance, as shown in Figure~\ref{fig:scatter_time_rq2}, there are project versions with nearly $4k$ test cases from the $Time$ project, which have less preparation time than project versions with around $3k$ test case from the $Math$ project. This is due to the fact that the average test code for the $Math$ project is longer than that for the $Time$ project, resulting in its AST tree size being twice as long. Therefore, although the test suite size of the $Math$ project is smaller, its preparation time is significantly longer than that of the $Time$ project. In contrast, the preparation time for \ourapproach~is not affected by the length of the test code, which further supports our claims regarding its scalability.

\vspace{2pt}
\textbf{\textit{Search Time.}}
We observe that the time \ourapproach~takes for searching the optimal test suite (mean~=~$0.59$ min and median~=~$0.21$ min across projects) is also shorter than that for ATM/Combined (mean~=~$0.91$ min and median~=~$0.29$ min) and ATM/TreeEditDistance (mean~=~$0.64$ min and median~=~$0.23$).

Moreover, quadratic regression reveals that, while quadratically increasing with the number of test cases per version, the search time of \ourapproach~increases more slowly with the squared number of test cases per version (a coefficient on the quadratic term equal to $7.013e^{-06}$) than ATM/TreeEditDistance ($7.844e^{-06}$) and one order of magnitude more slowly than ATM/Combined ($2.680e^{-05}$), with more than $79\%$ ($R^2$) of the variation in \textit{MT} being explained by the number of test cases.

Note that the search time not only depends on the number of test cases per version but is also impacted by the number of generations required by the GA to converge. For instance, as shown in Figure~\ref{fig:scatter_time_rq2}, though the number of test cases per version for the $Math$ project ($2,078$) is less than that for the $Time$ project ($3,918$), GA needed more generations to converge for $Math$, thus resulting in longer search time. Test cases of the $Math$ project are more diverse than that of the $Time$ project, with a higher standard deviation of similarity values ($0.13$ compared to $0.09$), thus leading to more generations and longer convergence time for the GA search.

In conclusion, in addition to being more efficient than ATM in terms of preparation time, the similarity measure employed by \ourapproach~provides better guidance and thus helps the search algorithm converge faster than ATM, thus resulting in less search time. 

\summarybox{\textbf{RQ2 summary.}
\ourapproach~outperforms ATM by achieving significantly higher \textit{FDR} ($0.84$ versus $0.81$, on average). However, the main benefit of \ourapproach~is that it is running much faster than ATM ($0.82$ min versus $4.06$ min, on average), in terms of both preparation time (up to two orders of magnitude faster) and search time (up to one order of magnitude faster). Given the acute scalability issues met in industrial contexts regarding minimization for large test suites and systems, this result is of high practical importance.}

\subsection{Discussion}
\label{Discussion}
\subsubsection{Scalable test suite minimization with much less preparation time than the SOTA.}
Our results show that \ourapproach~runs up to two orders of magnitude faster than the SOTA approach (i.e., ATM) while achieving a statistically significantly higher fault detection rate.
We should note that, according to the average \textit{MT} across projects, \ourapproach~runs five times faster than ATM/Combined and $44$ times faster than ATM/TreeEditDistance. These differences tend to increase significantly as the test suite size grows. For example, for the $Math$ project with $2078$ test cases, \ourapproach~runs $6$ and $102$ times faster than ATM/Combined and ATM/TreeEditDistance, respectively. This suggests that \ourapproach, while achieving higher effectiveness, can save enormous time and resources for minimizing test suites of the much larger software systems and test suites encountered in practice.
The preparation phase plays a crucial role in the scalability of \ourapproach. Since we employed pre-trained language models to extract vector-based test method embeddings and used a highly optimized function to compute similarity between them, we managed to achieve much shorter preparation times compared to ATM. 
In addition, since we used the test code as input without preprocessing it and converting it into other formats, such as ASTs for ATM, we saved substantial memory space as well. 

Unlike coverage-based techniques, the size of a project (i.e., source lines of code) has no impact on the minimization time of black-box TSM approaches, including {\ourapproach} and ATM, as minimization time mostly depends on the number of test cases per project version. In addition, we optimize the search process by changing the data structure for input similarity values and accelerating the fitness calculation, which reduced the search time by $190.48$ times. The $Closure$ project---the largest by far, comprising $179$ KLoC and $174$ versions, with $7,308$ test cases per version---took only $17.80$ min in terms of \textit{MT} and achieved a \textit{FDR} of $0.79$, for a 50\% minimization budget. Note that for ATM, $Closure$ was far too large to even run our experiments, thus further illustrating the scalability gains with \ourapproach.

Though {\ourapproach} uses the same search termination criterion as ATM, which is defined as a fitness improvement of less than $0.0025$ across generations, we observed that \textit{FDR} reaches a plateau after a certain number of generations (e.g., about an average of $65$ generations across project versions using \ourapproach~with UniXcoder/Cosine). However, the GA search continues as long as there is sufficient improvement in fitness (e.g., \ourapproach~with UniXcoder/Cosine reached a maximum of $100$ generations). This suggests that the search termination criterion unnecessarily prolongs the search and thus needs further investigation. For example, using a fixed number of generations, determined based on experience with past versions, could save time and resources while achieving high \textit{FDR} values.

\subsubsection{Effective test suite minimization reduces testing cost.} 
Further, {\ourapproach} achieves an average saving of $41.72\%$ in test execution time with a $50\%$ test suite minimization budget. However, though this is due to the relatively modest size of the systems we experiment with, one might point out that the average execution time of test suites in our dataset is smaller than {\ourapproach} minimization time. Nevertheless, {\ourapproach} still offers significant advantages as testing is performed frequently, especially in CI contexts, while minimization is occasional, unlike test case selection and prioritization. It can thus achieve substantial savings in time and resources over time, particularly in CI environments, where numerous builds are tested for every code change, thus warranting the elimination of redundant or unnecessary test cases to significantly reduce the overall test execution time and required testing resources (e.g., memory, CPU).  

\subsubsection{More informative similarity measures that better guide the search.}
Our results show that the similarity measures \ourapproach~employed better guide the GA to converge faster to a higher \textit{FDR}, when compared to ATM, confirming the usefulness of using pre-trained language models for extracting test method embeddings.

Results also suggest that the \textit{FDR} results are influenced by multiple factors including the language models, distance functions, and test code characteristics.
Language models differ in the types of input used for pre-training, the model architectures employed, and the kinds of tasks they were pre-trained for, thus generating embeddings with various characteristics.
While Cosine Similarity captures directional similarity by measuring the angle between embeddings~\cite{sirisha2019cosine}, Euclidean Distance captures magnitude similarity by measuring the straight-line distance between embeddings in a multidimensional space~\cite{deza2009encyclopedia}.
Therefore, using different combinations of language models and distance functions is expected to yield varying \textit{FDR} results.

In addition, the test code characteristic (e.g., domain-specific character strings or complex string manipulations) of certain projects may also impact \textit{FDR} performance, depending on the input (test code or ASTs) and dataset (size and diversity of programming languages ) used for pre-training the language models.
In conclusion, the performance of \ourapproach~in terms of \textit{FDR} may be affected by several confounding factors. Future research should explore the effects of these factors through deeper analyses of test code structures and controlled experiments.

\vspace{3pt}
\subsubsection{Language models with fewer parameters may offer a better trade-off than those with many more parameters in TSM contexts. 
}
The results show that UniXcoder ($125 Million$ parameter size) outperforms CodeLlama-7b ($7 Billion$ parameter size) in terms of both effectiveness and efficiency. This suggests that larger language models, though comprising substantially more parameters and being pre-trained on larger datasets, may not necessarily perform better for certain downstream tasks such as TSM. 
CodeLlama-7b, as a decoder-only model, was pre-trained for code generation tasks, such as code infilling. While generating (infilling) the next token, the structure of the decoder makes each token only learn information from preceding tokens, thereby potentially making its embedding less informative due to the absence of information from subsequent tokens. In contrast, code understanding tasks (e.g., MLM) allow the model to learn the information before and after each token, thus making it more effective in learning the contextual semantic information from the source code. Furthermore, UniXcoder enhanced the code representation by utilizing AST as input, which further enables the model to understand the source code structure and grammar, thus resulting in better \textit{FDR}.

Note that, to the best of our knowledge, there is no publicly available open-source language model that outperforms UniXcoder on the code clone detection task, which is similar in some ways to our minimization task. More importantly, given that scalability is crucial for \ourapproach, larger language models may require much more memory, computational resources, and time when generating embeddings due to their substantially higher number of parameters and complex model architectures. Therefore, for a downstream task, such as TSM, that requires assessing test code similarity, relatively smaller language models can be a better option than larger ones.

\vspace{8pt}
\subsubsection{Using full code versus preprocessed code.}

We recommend that \ourapproach~be used with the source code of test cases without preprocessing, since it yielded better results than when using preprocessed code, as reported in Section~\ref{PreliminaryStudy} and also in Section~\ref{RQ2_results}. However, some aspects of preprocessing, such as normalizing variable identifiers, may be beneficial, since they do not have an impact on the data flow in the source code. Therefore, using such techniques may potentially help language models generate embeddings that are more generalizable and suitable for code similarity calculation across projects, thus improving the fault detection power of \ourapproach.

\section{Threats to validity}
\label{Threats}
\vspace{-1.5pt}

\vspace{2pt}
\noindent\textbf{Internal Validity}
Internal threats to validity are concerned with the ability to draw conclusions from our experimental results. To make fair comparisons and draw valid conclusions, we conducted our experiments using the same experimental design and dataset used to evaluate ATM~\cite{pan2023atm}. We evaluated the alternative \ourapproach~configurations and compared the best configuration to the two best configurations of ATM. For minimizing test suites, we used the same fitness function and parameter settings as ATM, but we further optimized GA by utilizing a more efficient data structure for storing the input similarity values and accelerating the fitness calculation. We compared the performance using optimized GA for both \ourapproach~and ATM. 

ATM used preprocessed code as input because of its very long preparation time. In contrast, {\ourapproach} used full source code, raising the question of whether differences in \textit{FDR} between \ourapproach~and ATM are due to differences in test representation (test method embeddings or ASTs) or code input (full code or preprocessed code).
To answer that question, we conducted an additional experiment to assess the best configuration of \ourapproach~(i.e., UniXcoder/Cosine) using preprocessed code, which showed that \ourapproach~with preprocessed code still fares better than ATM in terms of \textit{FDR}.

Another threat is related to the token length limit of language models, which might truncate the source code of some test cases~\cite{fatima2022flakify}, thus resulting in information loss that might in turn negatively impact the performance of \ourapproach. However, we observed that, for each version, the percentage of test methods with the token length exceeding the limit is only $2.98\%$ across projects, which can be considered negligible. Nevertheless, some projects have a relatively higher percentage of test methods exceeding that limit (e.g., $7.25\%$ for $Math$), which might explain the slightly lower \textit{FDR} of \ourapproach~compared to ATM on these projects. Future research should further investigate the use of additional language models with longer token length limits, though they might greatly impact efficiency.

\vspace{4pt}
\noindent\textbf{External Validity}
External threats are concerned with the ability to generalize our results. We assessed both the performance of ATM and \ourapproach~on a large dataset consisting of $16$ Java projects with a total of $661$ versions collected from \textsc{Defects4J} for which we could trace failures to faults. We further validated the performance of \ourapproach~using an additional larger project from Defects4J, called $Closure$, consisting of $174$ versions, for which we could not evaluate ATM. Though we did not evaluate \ourapproach~on projects written in other programming languages, \ourapproach~can be easily adapted to other programming languages as the language models we used were pre-trained on multiple programming languages and have been demonstrated to yield good performance in various downstream tasks in other programming languages~\cite{feng2020codebert,guo2020graphcodebert,guo2022unixcoder,lu2021codexglue}. Future research should extend our study to investigate \ourapproach~on projects with other programming languages to further generalize our conclusions.
In addition, test method execution time may vary from run to run. To measure the achieved time-saving in test execution time more precisely, we used the \textsc{Defects4J} infrastructure to run test suites $10$ times and collect the average execution time for each test method in a clean, stable environment, which ensures we run each test suite without interference and fluctuations in memory or CPU usage (e.g., exclusive access to a node). \footnote{We further collected the minimum and maximum execution times for each test method across the $10$ runs. The results show that the difference in \textit{TSR}, based on minimum and maximum execution times, ranges from 0.05\% to 1.38\% across all configurations, which indicates that the variation in test method execution times across runs is small and does not impact our conclusions regarding \textit{TSR}.}
However, we found many test cases ($24.14\%$ across projects) with $0$ execution time as they took less than $0.0001$ seconds to run. \footnote{To ensure this did not impact our results, we investigated the effect of these test cases on the Time Saving Rate by assuming their execution times to be $0.0001$---their maximum possible value---and re-calculating the Time Saving Rate. We found that the average Time Saving Rate of all configurations slightly increased by 0.08\% to 0.09\% only, which is negligible and thus does not invalidate our conclusions.} Future research should consider evaluating TSM approaches on much larger industrial systems.

\section{Conclusion}
\label{Conclusion}

In this paper, we proposed \ourapproach, a scalable and black-box similarity-based test suite minimization (TSM) approach based on pre-trained language models. We investigated five alternative language models---CodeBERT, GraphCodeBERT, UniXcoder, StarEncoder, and CodeLlama---to extract test method embeddings and two similarity measures (i.e., Cosine Similarity and Euclidean Distance) for similarity calculation based on these embeddings. We employed an optimized version of the Genetic Algorithm (GA) by utilizing an efficient data structure for handling similarity values to improve fitness calculation and find optimal subsets of test suites. We assessed the performance of alternative \ourapproach~configurations on a large dataset consisting of $17$ Java projects with a total of $835$ versions collected from \textsc{Defects4J}. We used the Fault Detection Rate (\textit{FDR}), Minimization Time (\textit{MT}), and Time Saving Rate (\textit{TSR}) as evaluation metrics to assess the effectiveness and efficiency of \ourapproach. We identified the best configuration of \ourapproach~and compared it to ATM, a recent black-box TSM approach that achieved better trade-off in terms of effectiveness and efficiency than the alternative SOTA TSM approach (FAST-R). Results indicate that the best \ourapproach~configuration~(i.e., UniXcoder/Cosine) outperformed the two best configurations of ATM by achieving significantly higher \textit{FDR} results ($0.84$ versus $0.81$, on average) and more importantly, running much faster ($0.82$ min versus $4.06$ min, on average) than ATM, in terms of both preparation time (up to two orders of magnitude faster) and search time (up to one order of magnitude faster), which is particularly important on much larger industrial systems and test suites, while achieving a saving of $41.72\%$ in test execution time for a $50\%$ test minimization budget.

\vspace{4pt}
\noindent\textbf{Future work.}
We plan to assess the performance of \ourapproach~in practice using large industrial software systems and projects in other programming languages to further generalize our conclusions. The main challenge is the ability to automatically trace test failures to system faults.

\section{Data Availability}
\label{DataAvailability}
The replication package of our experiments, including the data, code, results for other minimization budgets, and detailed \textit{FDR}, \textit{MT}, and \textit{TSR} results of \ourapproach~and baseline approaches, is available on Zenodo~\cite{our_replication_package}.

\bibliographystyle{unsrt}
\bibliography{reference.bib}

\begin{IEEEbiography}[
\vspace{-20pt}{\includegraphics[width=1in,height=1.25in,clip,keepaspectratio]{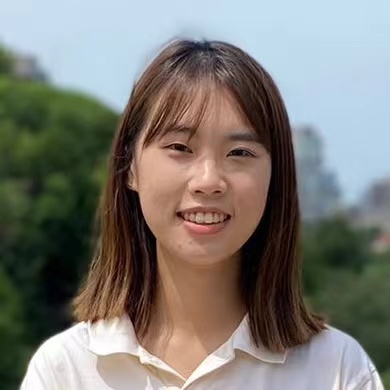}}]{Rongqi Pan}
  is currently working toward the PhD degree with the School of EECS, University of Ottawa and a member of Nanda Lab, Canada. She obtained her Master’s degree in Statistics at the University of Illinois at Urbana-Champaign, USA. Her research interests include automated software testing, natural language processing, and applied machine learning.
\end{IEEEbiography}

\begin{IEEEbiography}[
\vspace{-14pt}
{\includegraphics[width=1in,height=1.25in,clip,keepaspectratio]{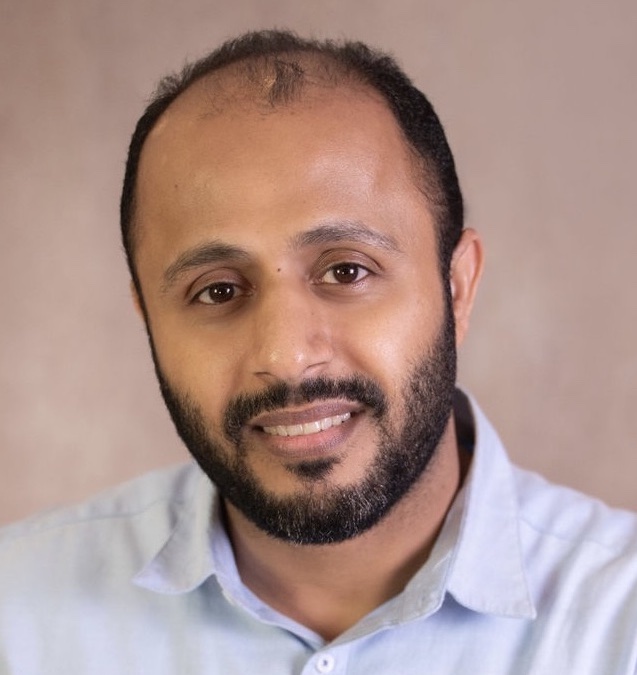}}]{Taher A. Ghaleb}
  is an Assistant Professor in the Computer Science Department at Trent University, Canada. He earned his Ph.D. in Computing from Queen’s University, Canada, where he was awarded the Ontario Trillium Scholarship, a highly prestigious award for doctoral students. Following his Ph.D., he worked as a senior research scientist at the University of Toronto and as a postdoctoral research fellow at the University of Ottawa. His research interests include continuous integration and delivery (CI/CD), data-driven software analytics, artificial intelligence for software engineering, and mining software repositories. For more information, visit \href{https://taher-ghaleb.github.io}{taher-ghaleb.github.io}.
\end{IEEEbiography}

\vspace{-13cm}
\begin{IEEEbiography}[
\vspace{-17pt}{\includegraphics[width=1in,height=1.6in,clip,keepaspectratio]{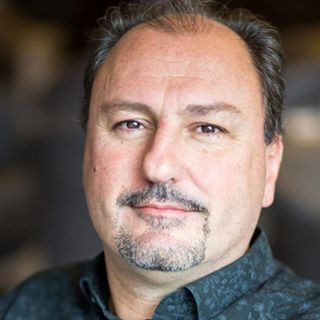}}]{Lionel C. Briand}
  is professor of software engineering and has shared appointments between (1) The University of Ottawa, Canada, and (2) The Lero SFI Centre---the national Irish centre for software research---hosted by the University of Limerick, Ireland. In collaboration with colleagues, for over 30 years, he has run many collaborative research projects with companies in the automotive, satellite, aerospace, energy, financial, and legal domains. Lionel has held various engineering, academic, and leading positions in seven countries.  He currently holds a Canada Research Chair (Tier 1) on "Intelligent Software Dependability and Compliance" and is the director of Lero, the national Irish centre for software research. Lionel was elevated to the grades of IEEE Fellow and ACM Fellow for his work on software testing and verification. Further, he was granted the IEEE Computer Society Harlan Mills award, the ACM SIGSOFT outstanding research award, and the IEEE Reliability Society engineer-of-the-year award. He also received an ERC Advanced grant in 2016 on modelling and testing cyber-physical systems, the most prestigious individual research award in the European Union and was elected a fellow of the Academy of Science, Royal Society of Canada in 2023. More details can be found at: http://www.lbriand.info.
\end{IEEEbiography}

\end{document}